\documentclass[aps,pre,floatfix,longbibliography,nofootinbib]{revtex4-2}
\usepackage{amsfonts}
\usepackage{amsmath}
\usepackage{amsthm}
\usepackage[x11names]{xcolor}
\usepackage{graphics}
\usepackage{graphicx,epstopdf}
\usepackage[export]{adjustbox}
\usepackage{subfigure}
\usepackage{amssymb}
\usepackage{lineno}
\usepackage[colorinlistoftodos]{todonotes}  
\usepackage[T1]{fontenc}
\usepackage{amsmath,amssymb,epsfig,latexsym,graphicx,dcolumn}
\usepackage{natbib}
\usepackage{todonotes}
\usepackage{svg}

\begin{document}
\setcounter{page}{1}
\title{Non-equilibrium wall model for large eddy simulations of \\ complex flows exhibiting turbulent smooth body separation }

\author{Rahul Agrawal}\email{rahul29@stanford.edu}
\affiliation{Center for Turbulence Research, Stanford University, }
\author{Sanjeeb T. Bose}
\affiliation{Cadence Design Systems, Inc. }
\affiliation{Institute for Computational and Mathematical Engineering, Stanford University,}
\author{Parviz Moin}
\affiliation{Center for Turbulence Research, Stanford University. }


\date{\today}

\begin{abstract}
In this work, a non-equilibrium wall model is proposed for the prediction of turbulent flows experiencing adverse pressure gradients, including separated flow regimes. The mean-flow non-equilibrium is identified by comparing two characteristic velocities: the friction velocity ($u_{\tau}$) and the viscous-pressure gradient velocity ($u_p$).  In regions where the 
pressure gradient velocity is comparable to the friction velocity ($u_p \sim u_{\tau}$), the near-wall turbulent 
closure is modified to include the effect of the pressure-gradient and convective terms. 
The performance of this wall model is evaluated in two canonical flows experiencing smooth body separation: the NASA/Boeing speed bump and the Bachalo-Johnson bump. Improvements in the predictive capabilities of the proposed model for the conventional equilibrium wall model are theorized and then demonstrated through numerical experiments.  In particular, the proposed wall model is able to capture 
the onset of boundary layer separation observed in experiments or DNS calculations at resolutions where the equilibrium wall model fails to separate.  


 \end{abstract}
\maketitle

\section{Introduction}
As the Reynolds number of a wall-bounded flow increases, the feasibility of direct numerical simulations (DNS) becomes impractical due to an increasing separation between the large, energetic scales of motion and the small, dissipative scales \cite{pope2000turbulent}. The large-eddy simulation (LES) paradigm is an approach in which the larger scales of motion of the flow are explicitly evaluated, while the effect of the unresolved scales on larger eddies is modeled using a ``subgrid-stress closure". Several models have been proposed, including both phenomenological models \cite{smagorinsky1963general,germano1991dynamic,nicoud_sigma,vreman1994formulation,rozema2015minimum,agrawal2022non} and data-driven methods \cite{chenyue2021artificial,zhou2019subgrid}. These approaches have been widely successful in predicting the correct turbulence statistics in canonical flows such as homogeneous isotropic turbulence \cite{nicoud_sigma}, turbulent channels \cite{bae2019}, and boundary layers \cite{yang2015integral}.  At low or moderate Reynolds numbers, it is computationally feasible for LES resolutions to resolve near-wall viscous length scales, and as a result, no-slip boundary conditions can be applied due to accurate estimates of the near-wall velocity gradients.  This is often referred to as a wall-resolved LES regime.\\

\noindent
For many engineering flows of interest, even wall-resolved large-eddy simulations are impractical since the grid-point scaling of wall-resolving LES is not too dissimilar from a direct numerical simulation \cite{choi2012grid}. In these cases, a ``wall model" is used to provide an estimate of the wall shear stress and heat flux in lieu of a no-slip closure
that cannot be accurately computed. The most commonly used wall model is the equilibrium wall model, which supplies the wall-shear stress to the outer resolved flow field, assuming
that thin boundary layer approximations are valid and a 
law of the wall is present in unresolved regions of the 
grid beneath the ``matching location.'' This model, originally conceived by \citet{deardorff1970numerical}, is based on the constant stress layer arguments \citep{cabot2000approximate} and has been successfully used beyond the canonical turbulent channel and zero-pressure gradient boundary layers in \emph{a-posteriori} LES. Recently, it was shown that wall-modeled LES with the equilibrium wall model correctly predicts the lift, pitching moment, and drag for a full aircraft system near stall conditions \cite{goc2021large,hlcrmkonrad}. 
Despite the fact that this model is formally invalid in the presence of strong pressure gradients, it has been 
reasoned that the equilibrium approximations are admissible 
if the ``matching location'' is sufficiently close to the wall \citep{bose2018wall}.  A tighter bound on the magnitude of these errors with resolution requirements that scale locally with $Re^{-2/3}$ has recently been established \citep{agrawal2023reynolds}.  The resulting local grid point scaling, $N \sim Re^{4/3}$, is more restrictive than prior estimates for wall-modeled LES, and in practice, these resolution requirements would be intractable for many practical flows at higher Reynolds numbers.\\

\noindent
There have been several non-equilibrium wall modeling approaches that have been introduced in order to relax the assumptions on the boundary layer introduced by equilibrium wall stress approximations \citep{wang2002dynamic,balaras1996two,hickel2013parametrized,park2014improved}. 
\citet{balaras1996two} included the unsteady, convective, and pressure gradient terms in their equations for the wall-parallel flow to solve the thin boundary layer equations, thus making a ``zonal model".  Wang and Moin \cite{wang2002dynamic} then proposed a model without including the convective terms. Later, Hickel \cite{hickel2013parametrized} showed that the pressure gradient and the convective terms play a similar role and should be included together. Importantly, all these models utilized an equilibrium closure for the Reynolds shear stresses in their thin-boundary layer equations. \citet{park2014improved,park2016numerical} developed an unsteady 3D RANS equation-based wall model where the eddy viscosity used to model the Reynolds stresses is computed dynamically by accounting for the effects of the resolved stresses on the skin friction. The requirement for the generation of a 3D embedded RANS mesh and interpolatory 
closures between the LES grid and the near-wall RANS mesh imparts significant complexity to the construction of the 
model and has limited its deployment in complex geometries.  Moreover, improvements offered by such an approach depend on the accuracy of RANS models in 
adverse pressure gradient regimes, including separation.  While the coupling with the interior LES can reduce the RANS modeling burden, RANS models alone have struggled in predicting flows with separation \citep{slotnick2014cfd}. This suggests that the mere inclusion of non-equilibrium effects into the wall model does not necessarily lead to more accurate solutions. \citet{kamogawa2023ordinary} formulated the first non-equilibrium-based wall model that is completely local in the wall-parallel plane while accounting for both the convective and the pressure gradient terms. However, the authors observed marginally worse predictions of the wall-shear stress compared to the standard equilibrium wall model for a boundary layer with a blowing and suction-based top wall to induce flow separation as in Na and Moin \citep{na1998direct}.  \\

\noindent
Thus, there remains a need for improvements in-wall models for LES in flows experiencing boundary layer separation. There have also been observations of non-monotonic convergence upon grid refinement in the prediction of the size of turbulent separation bubbles depending on the choices of subgrid-scale and wall models \citep{gocsubgrid,goc2020wall,agrawal2022non}.  
Accurate solutions can be obtained, but the required grid resolutions may engender prohibitive resolution requirements in more complex flow settings.  The Reynolds number scaling of the wall stress errors advanced in 
\citet{agrawal2023reynolds} suggests the 
the fundamental issue remains the incorporation of the effects of imposed pressure gradients.  The theoretical analysis suggests that these errors are significant when the velocity scale imposed by the pressure gradient 
($u_p^3 = \frac{\nu}{\rho}\frac{dP}{dx}$) is larger than the near-wall friction velocity ($u_{\tau}$)
, and when the LES grid resolution ($\Delta$) is resolved with respect to friction velocity length scale ($\Delta u_{\tau}/\nu \sim O(1)$) but unresolved with respect to the pressure gradient ($\Delta u_p / \nu \gg 1$).  These conditions are experienced in the vicinity of a separation point at high Reynolds numbers.  
\\

\noindent
Finally, we also note that there are alternative modeling strategies for the modeling of the unresolved near-wall effects that do not make a direct appeal to the thin boundary layer approximation framework.  These include alternative formal boundary conditions 
\citep{bose2014dynamic,bae2019,whitmorebump} or purely data-driven approaches \citep{ling2022wall}. As this work investigates assumptions in the closure of the thin boundary layer equations, we do not make any direct comparisons against any of these approaches. \\

\noindent
The rest of this article is written as follows: Section II describes the governing equations of LES, the subgrid-scale model used in this work, and a brief discussion of the computational solver. Section III proposes a wall model
that includes both the pressure-gradient and convective terms. Section IV demonstrates the importance of capturing the effects of convective and pressure-gradient terms in wall-modeled LES of non-equilibrium flows in an \emph{a-priori} sense. Sections V and VI present \emph{a posteriori} 
validations of the proposed model on the Boeing speed bump and transonic Bachalo-Johnson bump.  Finally, conclusions are offered in Section VII. 

\section{Computational Methodology}
\subsection{Governing Equations}
\noindent
In this work, the filtered, large-scale field variables (such as velocity and pressure) in LES are denoted by $\overline{f}$, and their corresponding Favre average is denoted by $\widetilde{f}$.  The governing equations of the large-scale flow field are given by, 

\begin{equation}
\frac{\partial \overline{\rho} }{\partial t }
+ \frac{\partial (\overline{\rho} \; \tilde{u}_i )}{\partial x_i} = 0  
\end{equation}

\begin{equation}
\frac{\partial  (\overline{\rho} \; \tilde{u}_i )}{\partial t }+\frac{\partial (\overline{\rho} \; \tilde{u}_j \;  \tilde{u}_i )}{\partial x_j } =- \frac{\partial \overline{p}}{\partial x_i } + \frac{\partial (\mu \tilde{S^d}_{ij} ) }{\partial x_j } -\frac{\partial \tau^{sgs}_{ij}}{\partial x_j} ,
\end{equation}

and

\begin{equation}
\frac{\partial  \overline{E}}{\partial t }+\frac{\partial (\overline{E} \; \tilde{u}_j)   }{\partial x_j } =- \frac{\partial (\overline{p}\; \tilde{u_i}) }{\partial x_i } + \frac{\partial   (\mu \tilde{S^d_{ij}} \tilde{u_i} )}{\partial x_j } -\frac{\partial (\tau^{sgs}_{ij} \tilde{u_i})  }{\partial x_j} - \frac{\partial Q_j^{sgs} }{\partial x_j } + \frac{ \partial }{\partial x_j } ( \kappa \frac{\partial \overline{T}}{\partial x_j }) ,
\end{equation}
where,
$\overline{E} =\overline{\rho} \; \overline{e} + 0.5 \;  \overline{\rho}\tilde{u_i}\tilde{u_i}$ is the sum of the resolved internal, $e$,  and kinetic energies,  $\rho$ is the density,   $T$, is temperature, $\mu(T)$ is viscosity, thermal conductivity is $\kappa(T)$ and velocity vector $\vec{u} \; = \; \{ u_1, \;u_2, \;u_3\}$. $\tilde{S^d}_{ij}$ is the deviatoric part of the resolved strain-rate tensor. The relationship between the temperature and the molecular viscosity is assumed to follow a power law with an exponent of 0.75. A constant molecular Prandtl number approximation ($Pr = 0.7$) is used to obtain the thermal conductivity. Two additional terms, $\tau^{sgs}_{ij}$ and $Q^{sgs}_{j}$ require modeling closure. The subgrid stress tensor, $\tau^{sgs}_{ij}$ is defined as  $\tau^{sgs}_{ij} = \overline{\rho } (\widetilde{u_i u_j} - \tilde{u}_j   \tilde{u}_i) $. Similarly, $Q_j^{sgs} = \overline{\rho} (\widetilde{e u_j} - \tilde{e} \tilde{u_j} ) $ is the subgrid heat flux. The dynamic, tensorial coefficient subgrid-scale model \citep{agrawal2022non} is used in the entirety of this work to model the unresolved scales of motion. The subgrid heat flux is modeled using the constant turbulent Prandtl number approximation ($Pr_t = 0.9$).  


\subsection{Numerical solver and gridding practices}
The simulations presented in this article were performed using an explicit, unstructured, finite-volume solver, charLES, which solves the compressible Navier-Stokes equations as described in the preceding subsection. This code is 2\textsuperscript{nd}-order accurate in space, and 3\textsuperscript{rd}-order accurate in time, and utilizes Voronoi diagram-based grids. More details of the solver and validation cases can be found in \citet{bres2018large,goc2021large,hlcrmkonrad,agrawalarb2023_2}. Formally skew-symmetric operators on unstructured grids are derived and used to conserve kinetic energy in the inviscid, zero Mach number limit in the current simulations. The numerical discretization also approximately preserves entropy in the inviscid, adiabatic limit. The computational grids employed in this work are based on a Voronoi diagram generated from a hexagonally close-packed lattice. 
In this work, control volumes are refined isotropically by factors of two in the layers near the boundaries during grid refinement studies. 

\section{Proposed near-wall model}
\noindent
In this section, a new wall model is proposed to improve the prediction of mean wall stress in the vicinity and inside of a turbulent separation bubble.

\subsection{Near-wall velocity scalings in flows with pressure gradients}
\noindent
Consider a turbulent boundary layer with an imposed mean pressure gradient. \citet{tennekes1972first} and separately, \citet{simpson1983model} suggested that the two characteristic viscosity-driven velocity scales that may govern the near-wall dynamics of a wall-bounded flow experiencing pressure gradients are the shear velocity, $\vec{u_{\tau}}$ and pressure-gradient based velocity, $\vec{u_p}$. \citet{simpson1983model} also showed that the $|\vec{u_p}|$ scaling is representative of the ``backflow velocity" in the vicinity and inside the separation bubble aft of the turbulent flow over a backward-facing step. \citet{stratford1959prediction} used mixing layer arguments to hypothesize a square-root power law velocity scaling (in terms of $u_p$) in the inner region of a flow at the point of separation. In the absence of a pressure gradient, the flow can be considered to be ``under-equilibrium", or the friction velocity governs the near-wall dynamics. Similarly, the other limit is obtained near a separation point when $u_\tau \rightarrow 0$ and the ``non-equilibrium" effects due to the pressure-gradient govern the near-wall dynamics \citep{stratford1959prediction}. The two velocity scales are defined as
\begin{equation}
     {u_{\tau, i}} = sign\left(U_i\right) 
     \left[\frac{\mu}{\rho} \frac{\partial |U_i| }{\partial y} \bigg|_{y=0}\right]^{1/2}   
\hspace{10pt}
\mathrm{and} 
 \hspace{10pt}
{u_{p,i}} =  sign \left(\frac{\partial P}{ \partial x_i}\right) \left[\frac{\nu}{\rho} \bigg|\frac{\partial P}{ \partial x_i}\bigg|\right]^{1/3}  
\label{eqn:utauupdef}
\end{equation}
where $y$ denotes the wall normal direction, and $i \; \in  \; \{1,2,3\}$ are the cardinal directions. Note that
repeated indicies in Equation \ref{eqn:utauupdef} does not imply summation. These two velocity scales can act along complementary directions or compete against each other, which may lead to four different conditions,
\begin{enumerate}
    \item Case I: $  \vec{u_p} \cdot \vec{u_{\tau}} > 0 ; \;  |\vec{u_p}| < |\vec{u_{\tau}}|  $
    \item Case II: $  \vec{u_p} \cdot \vec{u_{\tau}} > 0 ; \;  |\vec{u_p}| > |\vec{u_{\tau}}|  $
    \item Case III: $  \vec{u_p} \cdot \vec{u_{\tau}} < 0 ; \;  |\vec{u_p}| < |\vec{u_{\tau}}|  $
    \item Case IV: $  \vec{u_p} \cdot \vec{u_{\tau}} < 0 ; \;  |\vec{u_p}| > |\vec{u_{\tau}}|  $

\end{enumerate}


\noindent
For cases I and II, the pressure gradient is adverse with respect to the skin friction, and vice versa for cases III and IV. In cases I and III, the skin friction dominates the pressure gradient effects, and hence, the equilibrium conditions are presumably only weakly violated. However, in cases II and IV, the pressure gradient effects are dominant 
even in the viscous sublayer 
(as defined by $\frac{d}{dy}\left(\nu \frac{du}{dy}\right) = 0$), and a wall model that only relies upon these assumptions is not justifiable since the time scale set by the pressure gradient is faster and dominates the one from the skin friction. In the limit of a strong enough pressure gradient, Case II also represents a flow undergoing imminent separation. \\

\noindent
Next, we make estimates for a region where the wall closure would need to be modified from the standard equilibrium 
approximations even outside the viscous sublayer by comparing the $ |\vec{u_p}|  $ and $|\vec{u_{\tau}}| $ velocity scales. Under the statistically stationary conditions, the Lagrangian integral terms that appear in the integrated streamwise momentum equations are comprised of the combined effect of the convective and the pressure gradient terms, the integrated total shear term ( $\int^{x^+}_0 \tau^+_y dx^+ =  \int^{x^+}_0 (\frac{dU^+}{dy^+} - \overline{u^{+'} v^{+'} } )dx^+$), and the turbulent transport term ($ \int^{x^+}_0 \int^{y^+_{1} }_0  \frac{d \overline{u^{+'} u^{+'} } }{dx^{' +}}  dy'^+  dx'^+$ where $(\cdots)_{1}$ represents the value at the wall-normal location, $y_1$). In this analysis, it is assumed that the pressure gradient remains nearly constant from the wall up to the matching location ($\frac{dP}{dy}|_w \approx 0$) and the turbulent transport term can be ignored. \citet{wei2017integral} showed that even for boundary layers approaching separation, these two assumptions can lead to approximately correct description of the skin friction. The resulting approximate balance is given as follows, 
\begin{equation}
    \frac{ \int^{x^+}_0 \tau^+_y dx'^+ }{ \int^{x^+}_0  dx'^+}  \approx 1 +  \frac{ \int^{x^+}_0 \int^{y^+_{1} }_0  (\frac{ |\vec{u^3_p}|}{|\vec{u^3_{\tau}}|}  +  u^+ \frac{du^+}{dx^+} + v^+ \frac{du^+}{dy'^+} )  dy'^+  dx'^+ }{ \int^{x^+}_0  dx'^+} 
 \label{eqn:streamwiseintegral}
 \end{equation}
where the integrand in the denominator is $1$ because the system has been represented in the conventional inner viscous units (using $u_\tau, \nu$). The integral on the right side of this equation is referred to with the symbol, $\mathcal{H}$, in the remainder of this article. Equation \ref{eqn:streamwiseintegral} suggests that the constant stress layer assumption invoked in the equilibrium wall model may incur large errors when the second term on the right-hand side becomes large. It is, however, cautioned that this term becoming small only implies that the shear stress at an off-wall location and the wall stress balance in an integrated sense, not locally everywhere. Following \citet{kamogawa2023ordinary}, and replacing their model expression for the convective terms ($ u \frac{du}{dx} + v\frac{du}{dy} \approx - \frac{dP}{dx} \vert_{1} \frac{ u_{inner}^2}{U_{1}^2}$  and $ u_{inner}$ is the velocity profile below $y=y_1$), the right side of Equation \ref{eqn:streamwiseintegral} can be rearranged as, 

\begin{equation}
 \mathcal{H} \approx  \frac{ \int^{x^+}_0  \frac{ |\vec{u^3_p}|}{|\vec{u^3_{\tau}}|}  [ \int^{y^+_{1}}_0  dy^+  - \frac{ \int^{y^+_{match}}_0 u_{inner}^2  dy^+ }{U^2_{1}  } ]   dx^+ }{ \int^{x^+}_0  dx^+  } 
 \label{eqn:historymodelterm}
 \end{equation}

\begin{figure}[!ht]
    \centering
    {  
    \includegraphics[width=0.39\textwidth]{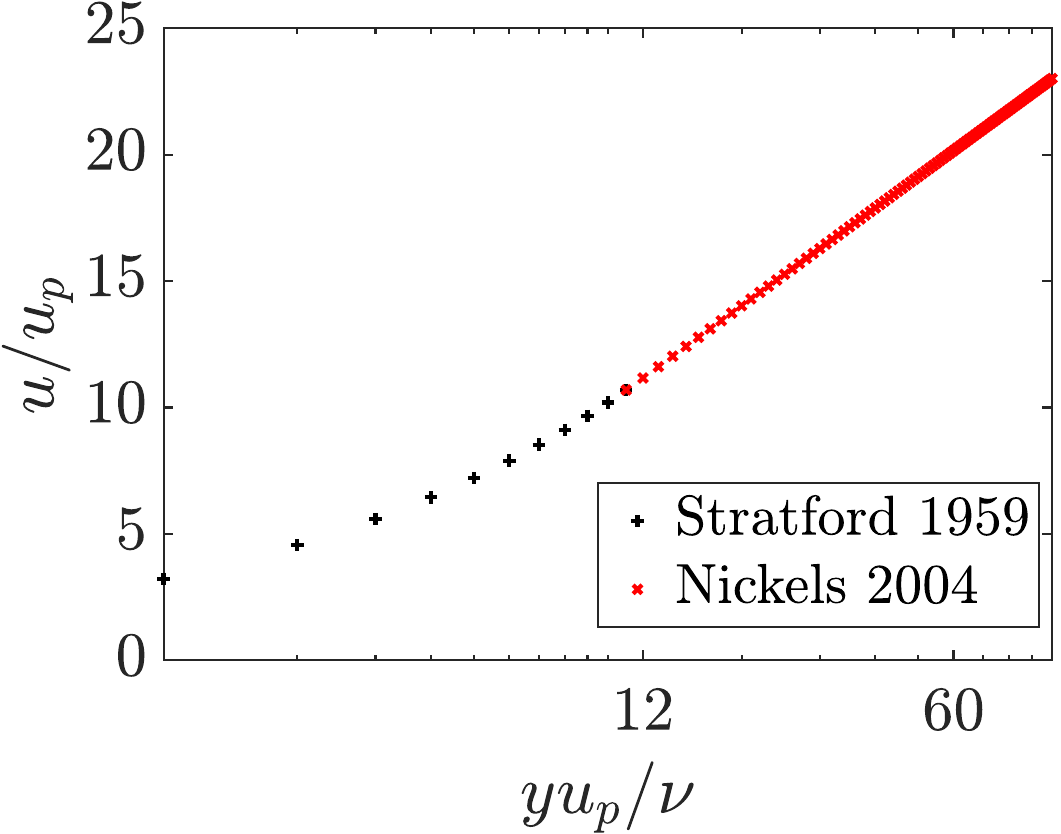}}
     \caption{ The wall-normal variation of the composite, $\mathcal{C}^{1}$ continuous, mean velocity profile used to close the integral of the convective terms in Equation \ref{eqn:eqwmsensorapprox5} near the separation point. The blending of the two velocity profiles occurs at $y u_p/\nu = 12$ per the viscously ``critically" stable Reynolds number arguments in \citet{nickels2004inner}.  }
     \label{fig:compositesepprofile}
 \end{figure}

\noindent
It is remarked that \citet{kamogawa2023ordinary} showed that their model form for the convective terms compares well with the DNS profiles of a flow over a flat plate with an imposed turbulent separation bubble (through blowing and suction boundary conditions) up to a $y/\delta \approx 0.1$ where $\delta$ is the local boundary layer thickness. For a boundary layer nearing separation, the inner mean velocity profile ($u_{inner}$) is assumed to be of a composite form (see Figure \ref{fig:compositesepprofile}), ensuring $\mathcal{C}^1$ continuity based on the near-wall profile per a ``half-power" law \citep{stratford1959prediction} and the logarithmic profile in the overlap-layer \citep{nickels2004inner}. \\ 

\noindent
Per \citet{agrawal2023reynolds}, a practically useful non-equilibrium wall model may warrant grid resolutions ($y_{1}$) of the order $y^p = \frac{u_p y_{1}}{\nu } \geq \mathcal{O}(10)$. In this limit, the numerator of Equation \ref{eqn:historymodelterm} can be simplified such that 
\begin{equation}
 \mathcal{H} \approx  \frac{ \int^{x^+}_0 \frac{ |\vec{u^3_p}|}{|\vec{u^3_{\tau}}|}  [ \int^{y^+_{1}}_0  dy^+  - \frac{ \int^{y^+_{1}}_0 u_{inner}^2  dy^+ }{U^2_{1}  } ]   dx^+ }{ \int^{x^+}_0  dx^+  } \approx   \frac{ \int^{x^+}_0 \frac{ |\vec{u^2_p}|}{|\vec{u^2_{\tau}}|} y_1^p [A log(y_1^p) + B]   dx^+ }{ \int^{x^+}_0   dx^+  }  \leq 
1\;  \mathrm{if} \;   \vert  y^p_1 [-A log(y_1^p) +B]  \frac{ |\vec{u^2_p}|}{|\vec{u^2_{\tau}}|}  \vert  \leq \mathcal{O}(1) 
 \label{eqn:historymodelterm2}
 \end{equation}
where $-A, \, B > 0 $ are some positive, real constants such that $-A log(y^p_1) + B$ is a good approximation of the wall-normal integration of the velocity profile, as used in Equation \ref{eqn:historymodelterm2}. Thus, $\mathcal{H}$ becomes non-negligible in the vicinity of a separation point when 
\begin{equation}
    \frac{ |\vec{u^2_p}|}{|\vec{u^2_{\tau}}|}  -\frac{\mathcal{O}(1)}{ y^p_1 [-A log(y_1^p) +B]  } \geq 0 
    \hspace{10pt} 
 \label{eqn:historymodelterm3}
 \end{equation}
The second term on the left side of the inequality is a decreasing function in $y^p_1$; we replace it with its maximum value within the range of $y^p_1$ of interest, $y^p_1 \gtrsim \mathcal{O}(10)$. Thus, 
\begin{equation}
    \frac{ |\vec{u^2_p}|}{|\vec{u^2_{\tau}}|}  - 1 \geq 0 
    \label{eqn:historymodelterm4}
 \end{equation}
is the simplified expression used to determine the strength of the pressure-gradient, even outside the viscous sublayer. In order to allow for 
standard equilibrium approximations to be utilized when the resolution is sufficient for the outer LES to capture
the effects of the pressure gradient (i.e., to deactivate the sensor when $y^p \leq \mathcal{O}(10)$), 
the inequality,

\begin{equation}
 \frac{y^p}{ 2 } \frac{ |\vec{u^2_p}|}{|\vec{u^2_{\tau}}|} -1 \geq 0     
    \label{eqn:historymodelterm5}
\end{equation}
is used (derived by replacing the near-wall, ``half-power" law profile of \citet{stratford1959prediction} in Equation \ref{eqn:historymodelterm2} instead of the composite profile in Figure \ref{fig:compositesepprofile}). Appendix I also discusses a connection of the proposed sensor with the critical stable viscous layer described in \citet{nickels2004inner}. 


\subsection{Wall Stress augmentation in strong non-equilibrium conditions }

\noindent
Once the sensor is deemed ``active" (i.e., that 
the pressure gradient effects are deemed significant), 
the equilibrium wall model's predicted stress is an inaccurate stress to feed to the outer flow since the conventional constant stress layer arguments are invalid. \citet{griffinincorporating} showed in an \emph{a-priori} sense that for non-equilibrium boundary layers, the stress experienced by the outer LES is different from the wall shear stress. Following this, we consider the steady Reynolds averaged boundary layer equations of LES as follows (for simplicity, the incompressible form of the governing equations is considered here),   
\begin{equation}
    \widehat{\widetilde{u}} \frac{   \partial \widehat{\widetilde{u}}  }{\partial x } +    \widehat{\widetilde{v}} \frac{  \partial \widehat{\widetilde{u}}   }{\partial y } \approx -  
\frac{1}{\rho} \frac{d \widehat{\widetilde{p}}}{ d x }   +  \nu \frac{\partial^2 \widehat{\widetilde{u}}   }{\partial y^2 } - \frac{  \partial \widehat{ \widetilde{u'}\widetilde{v'} }}{\partial y }   -  \frac{\partial \widehat{\tau^{sgs}_{12}}}{\partial y } 
\end{equation}
where $(\widehat{\cdots})$ is the Reynolds averaged value of the LES quantities (denoted by $(\widetilde{\cdots})$). Correspondingly, $\widehat{\widetilde{u}}$ and $\widehat{\widetilde{v}}$ are the streamwise and wall-normal mean LES velocities. The model of \citet{kamogawa2023ordinary}, as described in the preceding section is again used to approximate the convective terms. To perform the integral containing the pressure-gradient and convective terms, an approximate value of the pressure-gradient, $\frac{1}{\rho} \frac{d \widehat{\widetilde{p}}}{ d x }\vert_{match}$ is evaluated from a Taylor expansion from the quantities at the matching location.  Under the 
assumption that $\frac{1}{\rho} \frac{d \widehat{\widetilde{p}}}{ d y } \approx 0 $ at the wall, a Taylor series approximation for the pressure gradient is likely admissible.  Integrating the resulting system up to a matching location, $y_{match}$, with a matching velocity, $\widehat{\widetilde{U}}_{match}$, gives the approximate total stress as 
\begin{equation}
 \left( \nu \frac{d \widehat{\widetilde{u}}  }{dy}  -\widehat{ \widetilde{u'}\widetilde{v'} } - \widehat{\tau_{12}^{sgs}} \right) \bigg|_{y_{match}} \approx  \tau_w +   
\frac{1}{\rho} \frac{d \widehat{\widetilde{p}}}{ d x }\bigg|_{match} \; y_{match}  \left( 1 -  \frac{\int^{y_{match}}_0 \widehat{u}_{inner}^2 dy }{\widehat{\widetilde{U}}^2_{match} y_{match}}  \right)
   \label{eqn:eqwmsensorapprox5}
\end{equation}

\noindent
Thus, when the sensor is ``active", the right side of Equation \ref{eqn:eqwmsensorapprox5} is to be fed as the stress to the outer LES. Based on this equation, the key distinction in the present work compared to existing non-equilibrium wall models is when the sensor is ``active'', we attempt modeling the wall stress using the outer LES stress at the matching location, as opposed to invoking a RANS closure for it. This approach assumes that if the sensor activates in the vicinity of the separation point, the flow is nearly resolved with respect to the friction velocity length scales (as $y^+ \sim O(1)$). \citet{wang2002dynamic} advanced a similar argument to leverage the interior LES closures, but it was applied globally without regard for local flow conditions. It was found in this work that our approach increases the wall stress (compared to the equilibrium closure) locally, just before the onset of separation, thus reducing the velocity of the near-wall flow, which in turn supports flow separation of the outer LES. 
Finally, the ratio $\frac{\overline{u}_{inner}^2}{\overline{\widetilde{U}}^2_{match}}$ is modeled using the composite velocity profile constructed in Figure \ref{fig:compositesepprofile}. Further, consider the following application of the Cauchy-Schwartz inequality, 

\begin{equation}
1 - \frac{1}{\widehat{\widetilde{U}^2}_{match} y_{match}} \int^{y_{match}}_{0} \widehat{u}_{inner}^2 dy \leq 1 - \frac{1}{y^2_{match}} ( \int^{y_{match}}_0 \frac{\widehat{u}_{inner}}{\widehat{\widetilde{U}_{match}}} dy )^2
\end{equation}

\noindent
If an error (say $\chi \%$) is made in invoking a form of the mean velocity profile (under the assumption of a nearly uniform distribution of the error from $y = 0$ to  $y=y_{match}$), then the error in the model form of \citet{kamogawa2023ordinary} scales as $\chi^2 \%$ which is smaller than the error in the wall-stress ($\chi \%$). Thus, the choice of the model form for the convective terms is expected to produce smaller errors than the velocity profile used within the model form.

\subsection{Expected grid-point scaling with Reynolds number}

\noindent
Prior studies developing non-equilibrium wall models \citep{balaras1996two, wang2002dynamic, park2014improved, yang2015integral, griffinincorporating, kamogawa2023ordinary,bae2022scientific,ling2022wall} have
not estimated the required grid-point scaling to perform wall-modeled LES of separated turbulent flows.  Recently, \citet{agrawal2023reynolds} showed using a simplified Green's function solution near a separation point that flow length scales ``in equilibrium" for a separating boundary layer shrink with Reynolds number as $Re^{-2/3}$. Thus, for a computational code employing a nested-grid structure,  the total number of grid points ($N_{cv}$) for performing wall-modeled LES of turbulent boundary layers exhibiting smooth-body separation using an equilibrium wall closure scale as $N_{cv} \propto Re^{4/3}$.  \\

\noindent
Based on the experimental observations of \citet{jovic1995reynolds,adams1984experiments,devenport1991near} for the flow over a backward step, the peak of the skin-friction scales as $Re^{-1/2}$ in the separated flow region. Assuming this is a representative Reynolds number scaling in the vicinity of the separation point for other flows exhibiting a separation as $Re \rightarrow \infty$, $\mathcal{H}$ term in Equation \ref{eqn:historymodelterm2} scales as, 
\begin{equation}
   \mathrm{At \; the \; matching \; location,} \; \mathcal{H} \sim  {y^p}  \frac{ |\vec{u^2_p}|}{|\vec{u^2_{\tau}}|} \sim y Re^{\frac{1}{2}} \; \mathrm{as} \; Re \rightarrow \infty
\end{equation}
Thus, to capture the non-equilibrium component of the stress accurately independent of the Reynolds number, the grid resolution, $y_{match} \sim Re^{-1/2}$. This also recovers the linear grid-point scaling of wall-modeled LES with Reynolds number previously derived for flat-plate, equilibrium flows \citep{choi2012grid}. This expectation will be briefly tested in the \emph{a-posteriori} section of this article. 

\section{A PRIORI performance of proposed model}
\noindent
In the preceding section, a closure model including pressure-gradient and convective terms in wall-modeled LES of non-equilibrium flows was presented. In this section, we discuss in an \emph{a-priori} sense the importance of the pressure-gradient and convective terms, thus explaining the need to include the non-equilibrium component of the stress supplied to the outer flow. The error encountered in the stress supplied to the outer flow from the equilibrium wall model (which assumes a constant stress layer between the wall and the matching location) is compared with that from the proposed model (Equation \ref{eqn:eqwmsensorapprox5}) in an adverse pressure gradient flat-plate boundary layer \citep{bobke2017history} and over non-equilibrium airfoil flows \citep{vinuesa2018turbulent,tanarro2020effect}.  \\

\begin{figure}[!ht]
    \centering
    \includegraphics[width=1\textwidth]{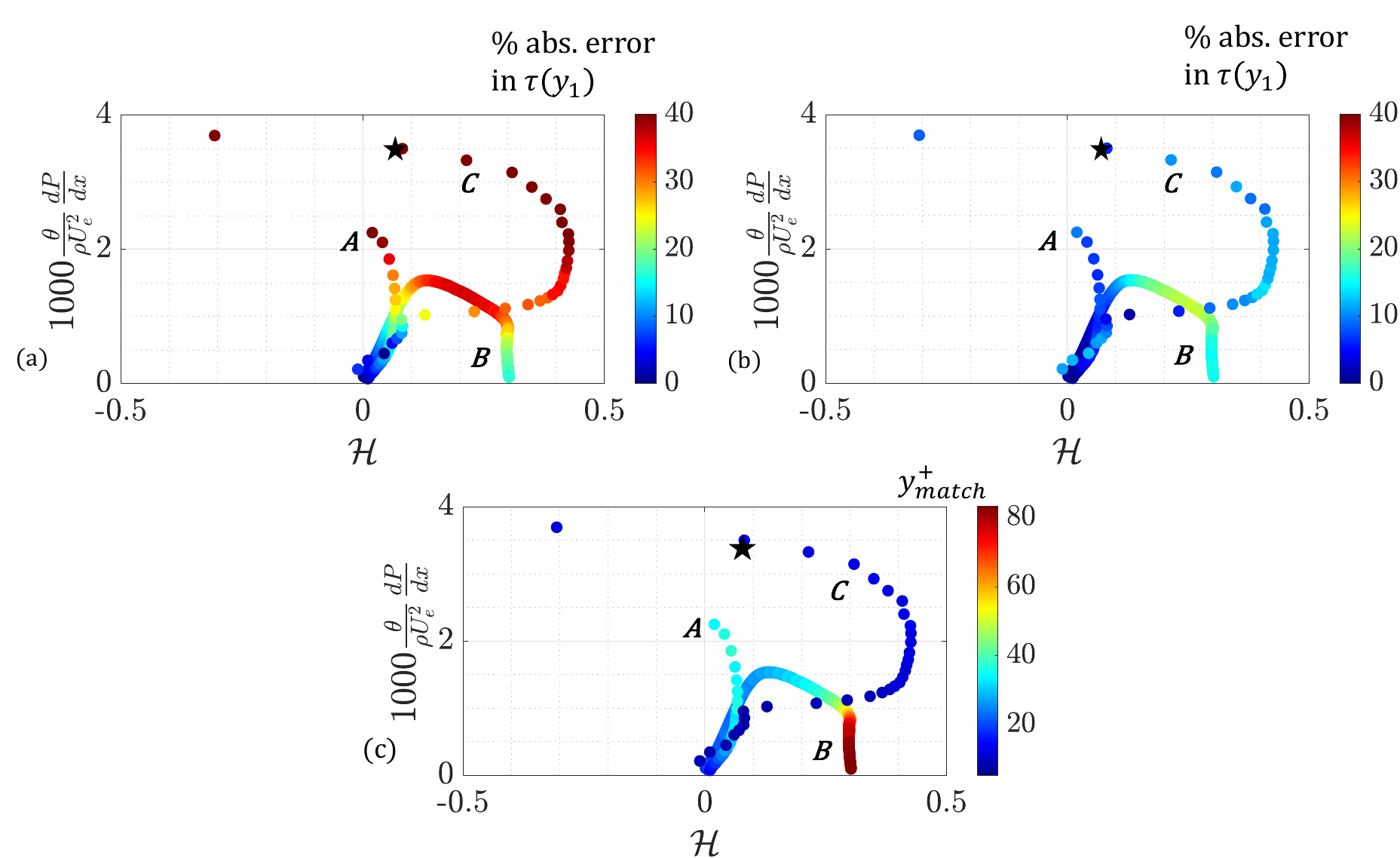}
    \caption{ \emph{A-priori} absolute percentage error distribution in the prediction of the total stress at the matching location as a function of $\mathcal{H}$ (see Equation \ref{eqn:historymodelterm2}) and the Alber parameter ($\theta/(\rho U^2_e) dP/dx $) for a fixed matching location in outer units, $y_{match}/\delta = 0.1$,  for (a) equilibrium wall model in \citet{lehmkuhl2018large} and (b) proposed model in Equation \ref{eqn:eqwmsensorapprox5}. Subfigure (c) denotes the height of the matching location in viscous units. The capital letters denote the data points from different flows: Data-series A corresponds to streamwise stations in flow over a NACA 0012 airfoil studied in \citet{tanarro2020effect}. Similarly, data-series B corresponds to a flat-plate boundary layer at $U_e \sim x^{-0.13} $ case studied in \citet{bobke2017history}. Data-series C corresponds to stations of flow over a NACA 4412 airfoil studied by \citet{vinuesa2018turbulent}. The starred point is a sample point chosen to aid the interpretability of this plot. This data point corresponds to NACA 4412 flow at $\Lambda \approx 0.0035, \mathcal{H} \approx 0.1$ at a matching location, $y^+_{match} \approx 13$. The percentage error in the shear-stress prediction at this point is $ 60 \%$ when employing the equilibrium wall model, and $12 \%$ when employing the proposed model.      }
    \label{fig:aprioriybydelta5}
\end{figure}

\noindent
Figure \ref{fig:aprioriybydelta5} shows the absolute errors in the prediction of the stress supplied to the outer solution at the matching location based on the equilibrium closure and the proposed closure for a fixed matching location, $y_{match}/\delta = 0.1$. The abscissa for this plot is $\mathcal{H}$ term from Equation \ref{eqn:historymodelterm2}, and the ordinate represents the flow-separation parameter of Alber \cite{alber1971similar}. This parameter is useful in identifying the outer-scale behavior of the boundary layer and also its proximity to flow separation (the threshold suggested by Alber \cite{alber1971similar} is $\Lambda = \theta/(\rho U^2_e) dP/dx = 0.004 $). It is apparent in Figure \ref{fig:aprioriybydelta5}(a) that the errors incurred by the equilibrium closure generally grow as $\mathcal{H}$ increases. However, as depicted in Cases A and C, the errors also exhibit a non-monotonic behavior with $\mathcal{H}$ as $\Lambda \rightarrow 0.004$ (intermittent flow separation). This implies that even at a constant outer resolution ($y_{match}/\delta =$ constant), the validity of the constant stress layer below the matching location can increase or decrease non-monotonically along the streamwise direction. On the contrary, in Figure \ref{fig:aprioriybydelta5}(b), the proposed wall model (in Equation \ref{eqn:eqwmsensorapprox5}) produces smaller errors and also does not exhibit the non-monotonic behavior observed with the equilibrium closure. Figure \ref{fig:aprioriybydelta5}(c) denotes the height of the matching location in viscous units; for cases A-C, as $\Lambda$ increases, the matching height decreases as the flow nears separation. Case B represents a flow with a freestream velocity profile governed by a power law ($U_e \sim x^{-0.13}$) studied by \citet{bobke2017history}; the proposed model produces errors that are smaller than the equilibrium closure.


\section{Wall-modeled LES of flow over a Gaussian bump}
\label{sec:bump}
\noindent
The development of CFD practices and models that predict smooth body separation was considered as a pacing item in the CFD 2030 Vision report \citep{slotnick2014cfd}. This section describes a canonical case exhibiting this phenomenon in the flow over the NASA/Boeing speed bump \citep{williams2020experimental,gray2021new,gray2022experimental,gray2022experimentalb}. In laboratory experiments, the solid body is shaped as a Gaussian in the streamwise direction and has tapered side-shoulders in the spanwise direction. The flow over the speed bump experiences a strong favorable pressure gradient followed by a strong adverse pressure gradient on the fore and aft sections of the bump, respectively. 
\citet{williams2020experimental} showed that at high Reynolds numbers, a turbulent separation bubble is formed on the aft side of the bump. Both the measurements of the skin friction and pressure drag in \citet{williams2020experimental,gray2022experimental} also suggested an approximate Reynolds number independence for $Re_L \geq 1.8 \times 10^6$. \\

\noindent
\citet{uzun2021high} performed a quasi-DNS (resolution of the outer part of the boundary layer resembles an LES) of this flow for a simplified, spanwise periodic variant of the geometry. Their results compare well with those from the experiments of \citet{williams2020experimental,gray2021new} at the midspan of the experimental configuration, and hence, the former is studied in this work. Within the context of wall-modeled LES,  \citet{zhou2023sensitivity,whitmorebump} have demonstrated the sensitivity of the prediction of flow separation to the choice of the subgrid-scale model, meshing topology, and the wall model. The ``building-block" machine-learning-based wall model of \citet{arranz2023wall} led to reasonable predictions of the separation bubble in this flow; however, they did not perform enough grid-refinements to exactly match the reference $C_p$ and $C_f$ values. \citet{agrawal2022non} used their non-Boussinesq subgrid-scale model with an equilibrium wall closure to produce accurate predictions of the separation bubble, but only on the most refined grids (finer than those simulated in \citet{arranz2023wall}). \\

\begin{figure}[!ht]
    \centering
    \subfigure[]{
    \includegraphics[width=0.7\textwidth]{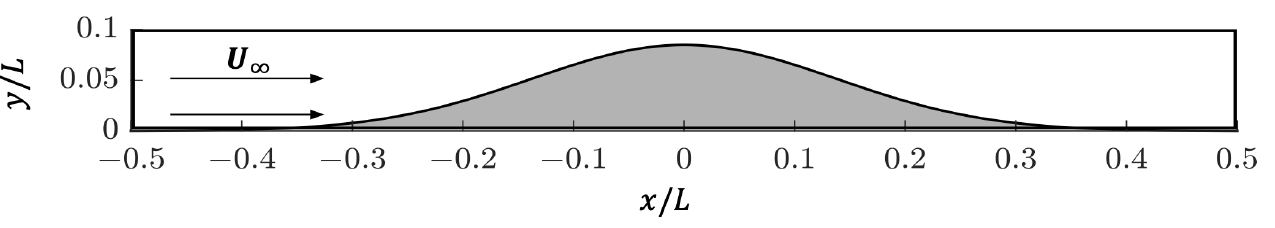}}
    \subfigure[]{
    \includegraphics[width=0.6\textwidth]{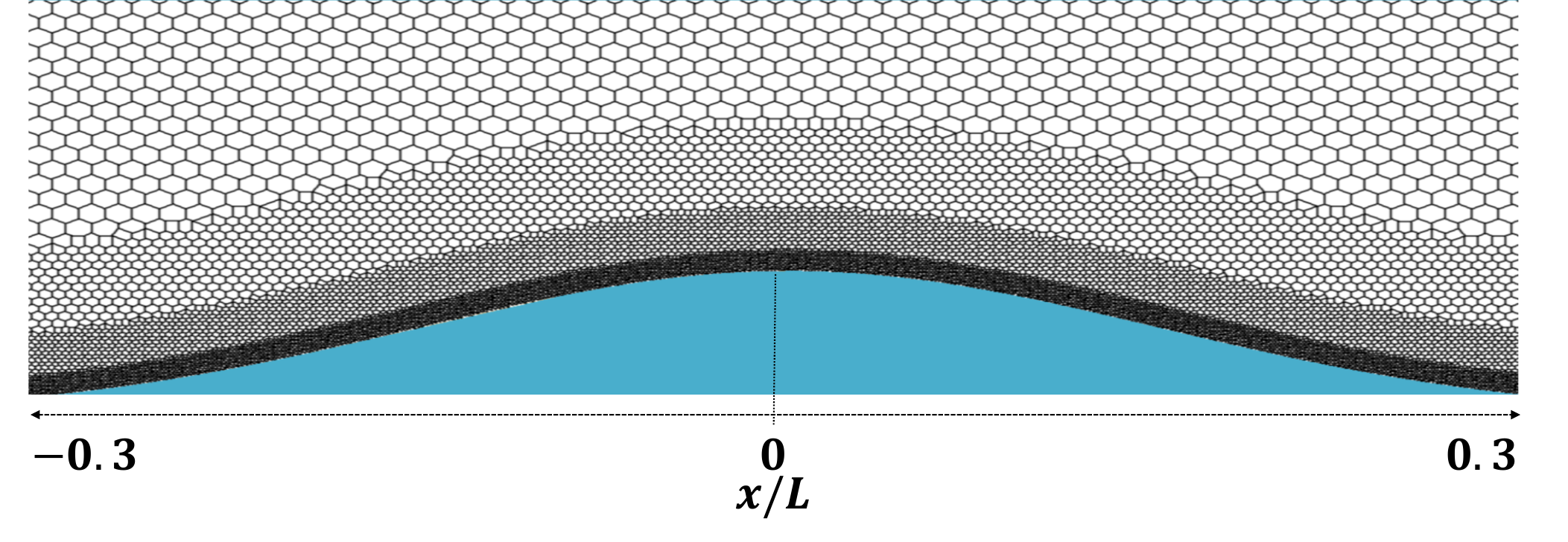}}
    \caption{ (a) A schematic of the simulation setup for the flow over the spanwise periodic NASA/Boeing speed bump at $Re_L = 2 \times 10^6$. The spanwise direction spans between $0 \leq z/L \leq 0.08$  and the top wall is at $y/L = 1.0$. (b) Cross section of the ``coarse" grid arrangement (details in Table \ref{table:resbump2}). Only the lower half of the vertical extent of the domain is shown. The top half is meshed uniformly up to the top wall. Three layers of isotropic refinement are visible adjacent to the bump surface. }
    \label{fig:2dbumpfig}
\end{figure}

\noindent
For this flow, the primary quantities of interest are the skin friction coefficient ($C_f$) and pressure coefficient ($C_p$) which are defined as 
\begin{equation}
    \frac{C_f}{2} = \frac{\tau_{w}}{\rho_{\infty} U^2_{\infty}} \hspace{5pt} \mathrm{and} \hspace{5pt}  \frac{C_p}{2} = \frac{ p - p_{\rm ref}}{ \rho_{\infty} U^2_{\infty}}.
\end{equation}
where $U_{\infty}$, $\tau_{w}$, $p$ and $p_{\rm ref}$ are the mean free-stream velocity, wall-stress, wall pressure, and the upstream reference pressure in the zero-pressure region of the flow respectively. Figure \ref{fig:2dbumpfig} provides a schematic of the simulation setup for this flow. 

\noindent
For the spanwise periodic case, the bump surface is defined by an analytical expression, $h(x,z)$, written as
\begin{equation}
    h(x) = \frac{h_0}{2} \, e^{-(x/x_0)^2}  ~,
    \label{eqn:bump}
\end{equation}
where $x$ is the streamwise direction and $h_0 = 0.085L$ is the maximum height of the bump; and $x_0 = 0.195L$ controls the Gaussian decay of the surface along the $x$ direction. The Reynolds number, $Re_L$ for this flow is defined in terms of the freestream velocity $U_{\infty}$ and the bump width, $L$. The computational grids and the grid-refinement practices are maintained to be the same as those in \citet{agrawal2022non,agrawalarb2022},  and some details are provided in Table \ref{table:resbump2}.\\

\begin{table}
\centering
\begin{tabular}{ p{1.5cm}p{1.5cm}p{1.5cm}p{1.5cm} }

Mesh & $N_{cv}$ & max $\Delta / L$  & min $\Delta / L$ \\
\hline\noalign{\vspace{3pt}}
Coarse  & $3$ Mil. & $1 \times 10^{-2} $ & $1.3 \times 10^{-3}$ \\
Medium  & $12$ Mil. & $1 \times 10^{-2} $ & $6.3 \times 10^{-4}$ \\
Fine    & $52$ Mil. & $1 \times 10^{-2} $ & $3.1 \times 10^{-4}$ \\
\end{tabular}
\caption{Mesh parameters for the quasi-DNS based spanwise-periodic case for the flow over the Boeing speed bump. }
\label{table:resbump2}
\end{table}

\noindent
In this work, we first study the $Re_L = 2 \times 10^6$ flow to match the aforementioned quasi-DNS \citep{uzun2021high}. Based on \citet{agrawal2023thwaites} who showed that the separation tendency of this flow is roughly independent of the exact inlet boundary condition, we feed a plug-flow at the inlet ($x/L = -1.0$). This leads to a finite flow development length (similar to the previous results of \citet{agrawal2022non,arranz2023wall,whitmorebump}). However, by $x/L = -0.6$ (well upstream of the bump), the skin friction agrees with the quasi-DNS. Free-stream conditions are set at the top boundary. A non-reflecting characteristic boundary condition with constant pressure is applied at the outlet (located at $x/L = 2.5$). The flux at the bump wall is provided using a wall model. 

\subsection{Predictions with proposed wall-model}

\begin{figure}[!ht]
    \centering
    \subfigure[]{
    \includegraphics[width=0.38\textwidth]{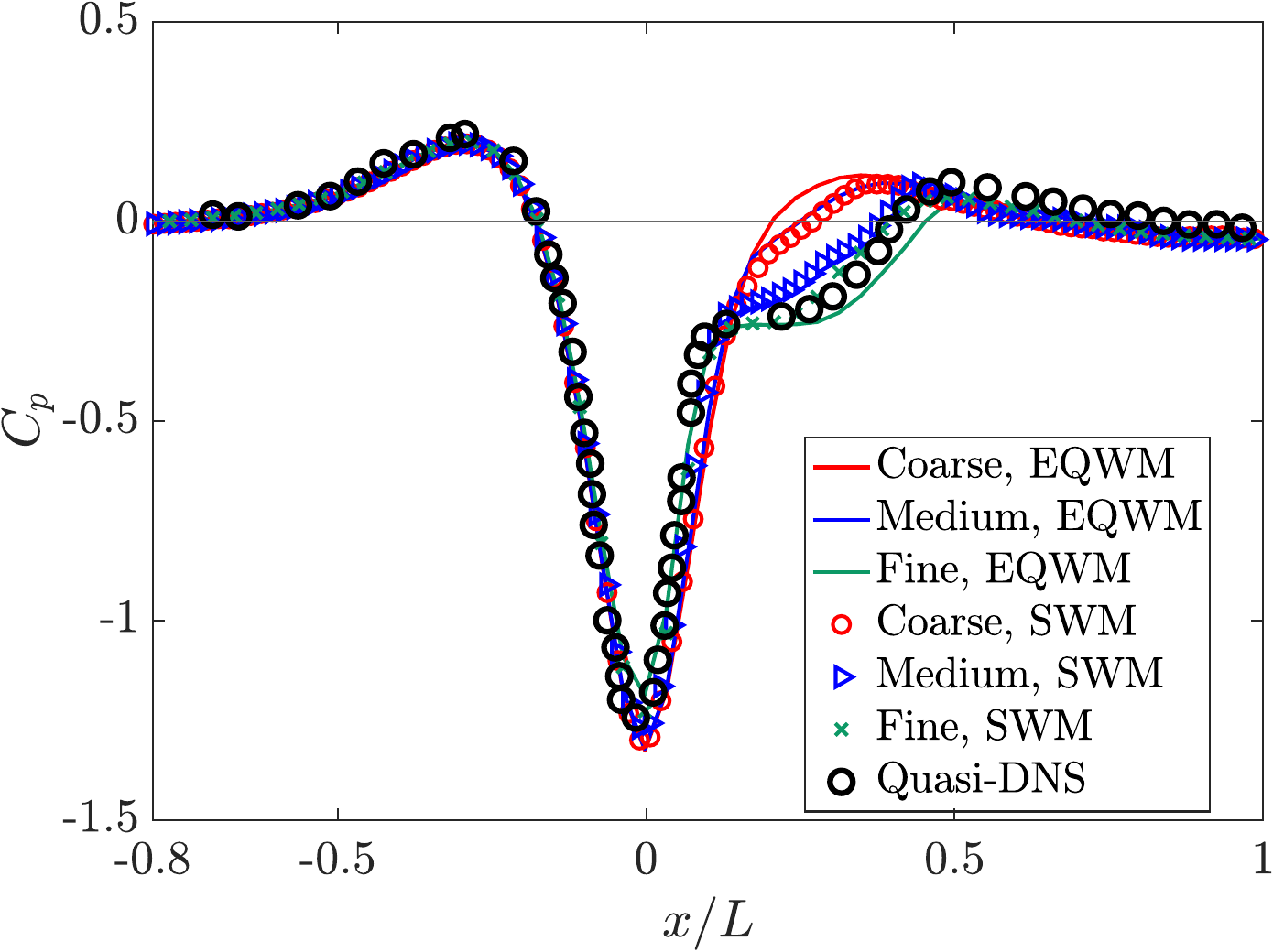}}
    \subfigure[]{
    \includegraphics[width=0.39\textwidth]{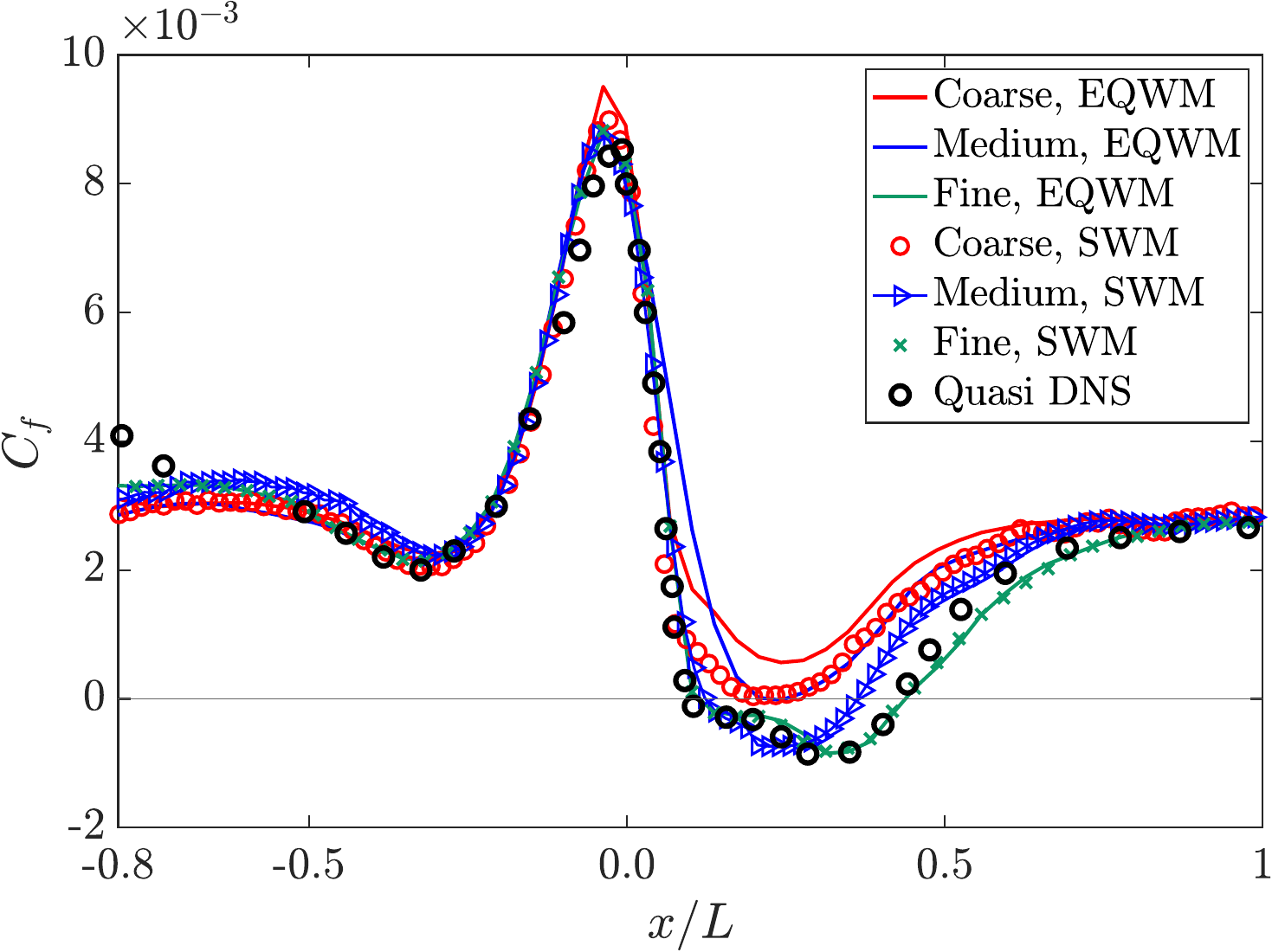}}
        \caption{  Streamwise distribution for the spanwise-periodic bump at $Re_L = 2 \times 10^6$ of (a) the surface pressure coefficient and  (b)the surface friction coefficient. The proposed non-equilibrium wall model is applied, and three mesh resolutions are shown. The black symbols represent the quasi-DNS of Uzun and Malik \cite{uzun2021high}. Here, SWM refers to the proposed sensor-based wall model in this work. The results corresponding to DTCSM/EQWM models were reproduced from \citet{agrawal2022non} with permission. Note that in subfigure (a), the blue solid line (medium grid, EQWM result) nearly overlaps with the red symbols (coarse grid, SWM result). This also holds in subfigure (b) for $x/L \gtrapprox 0.25 $ region.   }
    \label{fig:2dcfcp}
\end{figure}

\noindent
Figures \ref{fig:2dcfcp}(a) and \ref{fig:2dcfcp}(b) provide a comparison between the predictions of the surface pressure and the skin-friction from the standard equilibrium wall model \citep{lehmkuhl2018large} 
(denoted EQWM) and the proposed wall model (denoted SWM). This comparison is also made across a sequence of grid-refinements from ``coarse" to ``medium" and ``fine" (detailed in Table \ref{table:resbump2}). For both these figures, the predictions from the proposed wall model compare more favorably with the reference data than those from the equilibrium wall model. The equilibrium wall model predicts a flow separation on the finest grid only; however, with the proposed model, a separation bubble is predicted even on the medium grid (marked by the flattening of the $C_p$ curve or the occurrence of negative $C_f$). Secondly, the monotonic approach of WMLES with grid-refinement towards reference data remains preserved with the proposed model. In line with the results reported in \citet{agrawal2022non}, the skin-friction peak (which occurs when the inviscid pressure gradient flips its sign) is well captured for both wall models. Finally, the upstream skin friction between $-0.6 < x/L < -0.3$ is more accurate than those in other studies that employ similar inflow conditions but different wall-closures \citep{whitmorebump,zhou2023large,arranz2023wall}. Appendix II presents an \emph{a-posteriori} sensitivity of the prediction of wall-modeled LES to the choice of the order of magnitude of the second term in Equations \ref{eqn:historymodelterm3}-\ref{eqn:historymodelterm4}. \\ 

\begin{figure}[!ht]
    \centering
    \includegraphics[width=1.0\textwidth]{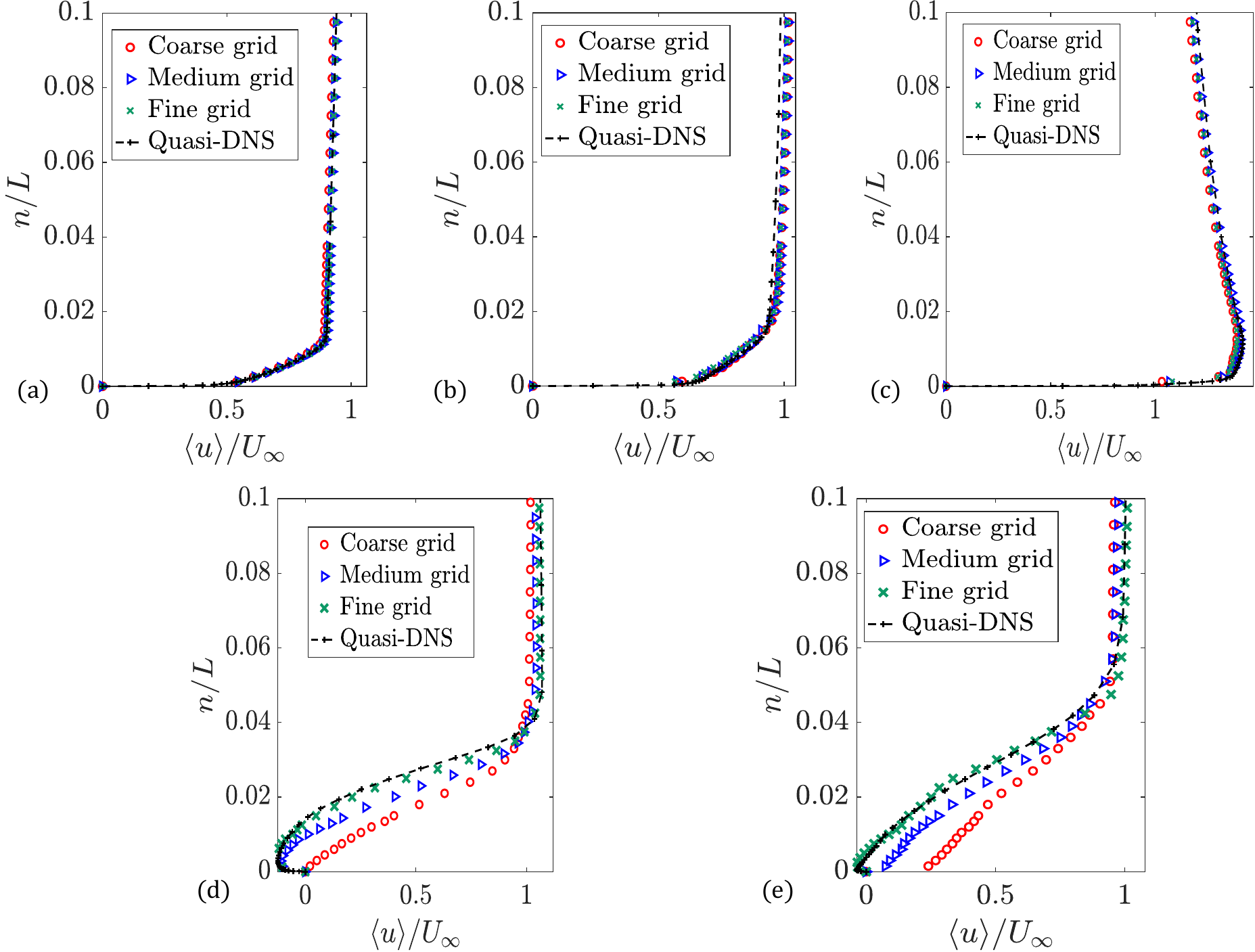}
    \caption{ The nondimensional streamwise velocity profiles, $\langle u \rangle / U_{\infty} $ at (a) $x/L=-0.4$, (b) $x/L=-0.2$, (c) $x/L=0.0$, (d) $x/L=0.2$,  and (e) $x/L=0.4$ for flow over the Boeing speed bump at $Re_L=2 \times 10^6$. The quasi-DNS refers to the study of Uzun and Malik \cite{uzun2021high}. The predictions are shown across the grid-refinement sweep for the tensorial subgrid-scale model and the proposed wall model combination.}
    \label{fig:2dvels}
\end{figure}



\noindent
The quality of the solutions is further probed by comparing the mean streamwise velocity profiles from wall-modeled LES to the quasi-DNS. In Figure \ref{fig:2dvels}, the profiles at five stations ($x/L = -0.4, \; -0.2, \; 0.0, \; 0.2, \; 0.4$) are plotted. For subfigures (a)-(c), which correspond to the pre-separation region, good and largely grid-insensitive agreement of the mean profiles with quasi-DNS is observed. In the post-separation region ($x/L \geq 0.1$, subfigures (d)-(e)), larger differences are observed with the agreement between wall-modeled LES and quasi-DNS improving as the grid is refined. The coarse and the medium grids predict a ``faster" velocity near the wall that is consistent with the reduced flow separation compared to the fine grid. \\

\begin{figure}[!ht]
    \centering
    \subfigure[]{
    \includegraphics[width=0.450\textwidth]{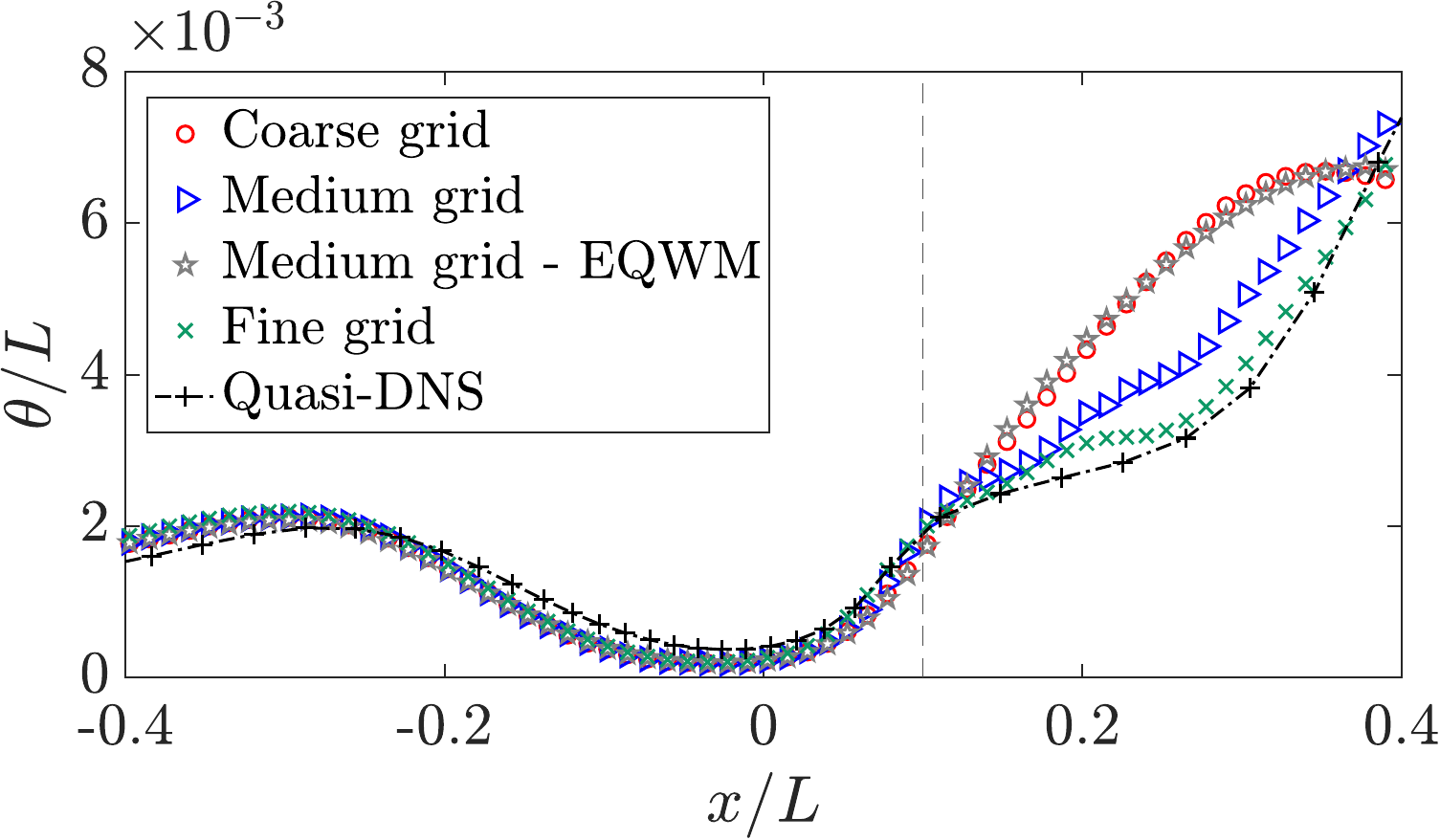}}
    \subfigure[]{
    \includegraphics[width=0.460\textwidth]{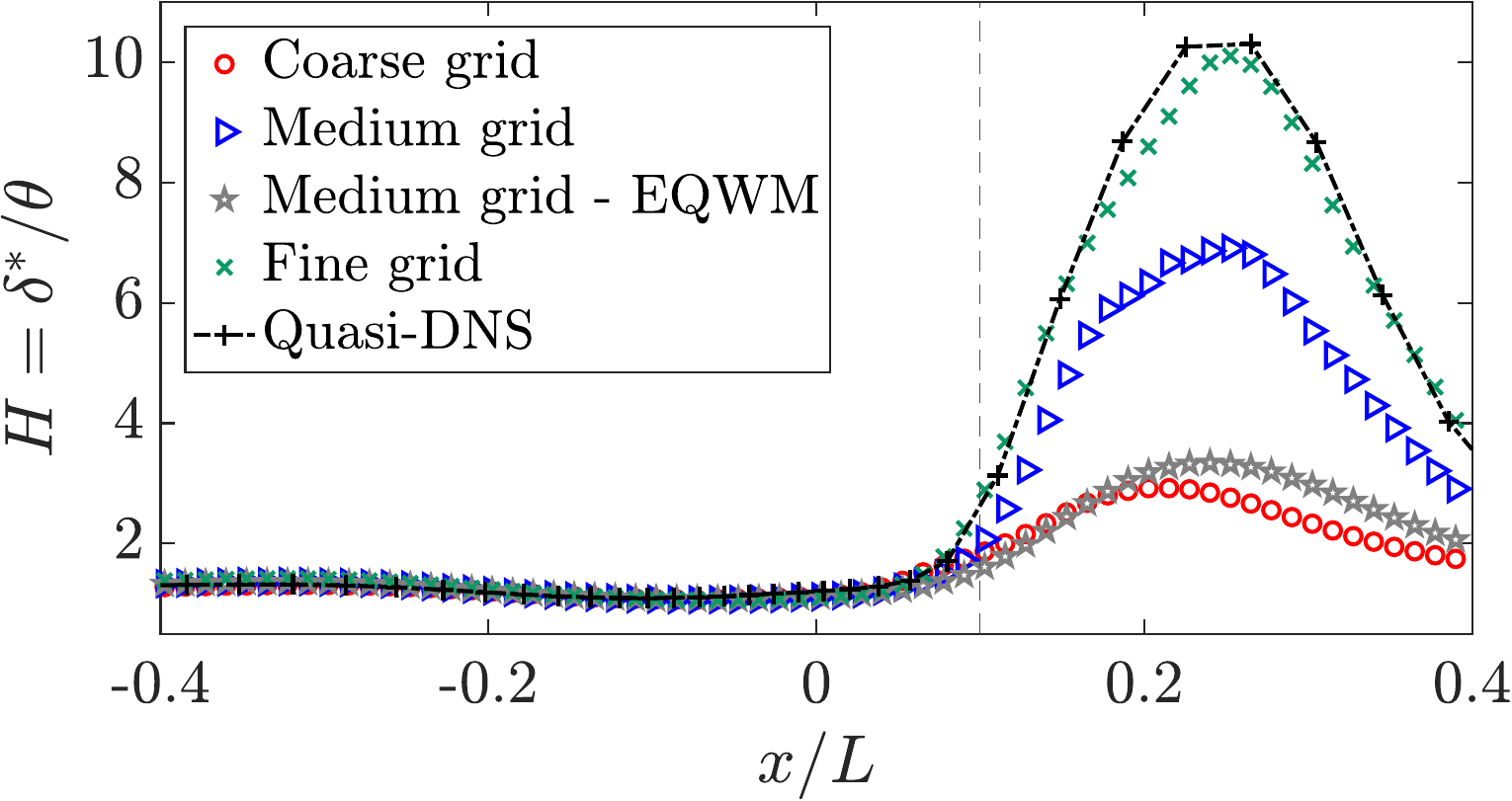}}
    \caption{ The predicted streamwise dependence of (a) the momentum thickness ($\theta$) and (b) the shape factor ($H=\delta^*/\theta$) across the grid-refinement sequence for the flow over the Boeing speed bump at $Re_L = 2 \times 10^6$. The quasi-DNS refers to the study of Uzun and Malik \cite{uzun2021high}. Unless otherwise stated, the proposed wall-model in this work is employed. The vertical dashed line denotes the point of separation in the reference quasi-DNS at $x/L=0.1$.   }
    \label{fig:2dshapetheta}
\end{figure}

\noindent 
Furthermore, the streamwise growth of the momentum thickness ($\theta$, refer to \citet{white2008fluid} for more details) and the shape factor ($H = \delta^*/\theta$) in these simulations is plotted in Figure \ref{fig:2dshapetheta}. The predicted wall-modeled LES solutions are in good agreement with the reference data for all the grids up to $\approx x/L = 0.05$, beyond which the occurrence (or absence) of flow separation determines the growth of the boundary layer. On the aft side of the bump, the attached boundary layers on the coarse grid and on the equilibrium wall model's medium grid lead to a nearly constant rate of growth of the boundary layer in the $0.1 \leq x/L \leq 0.3$ region. The adverse pressure gradient increases the shape factor up to $x/L \approx 0.25-0.30$, beyond which the pressure gradient on the attached flow becomes favorable. On the medium and fine grids employing the proposed model, $\theta$ qualitatively follows the diminished growth (due to a negative $C_f$ contribution in the Von-Karman integral equation). Similarly, the shape factor, $H$, increases rapidly beyond $x/L = 0.10$ (the mass deficit in the boundary layer rapidly increases in the separated region). The fine grid produces more favorable solutions consistent with its $C_p, \, C_f, \langle u \rangle$ being more accurate than the medium grid in the separated flow.

\noindent

\subsection{Activity of the proposed sensor}
\begin{figure}[!ht]
    \centering
    \includegraphics[width=0.5\textwidth]{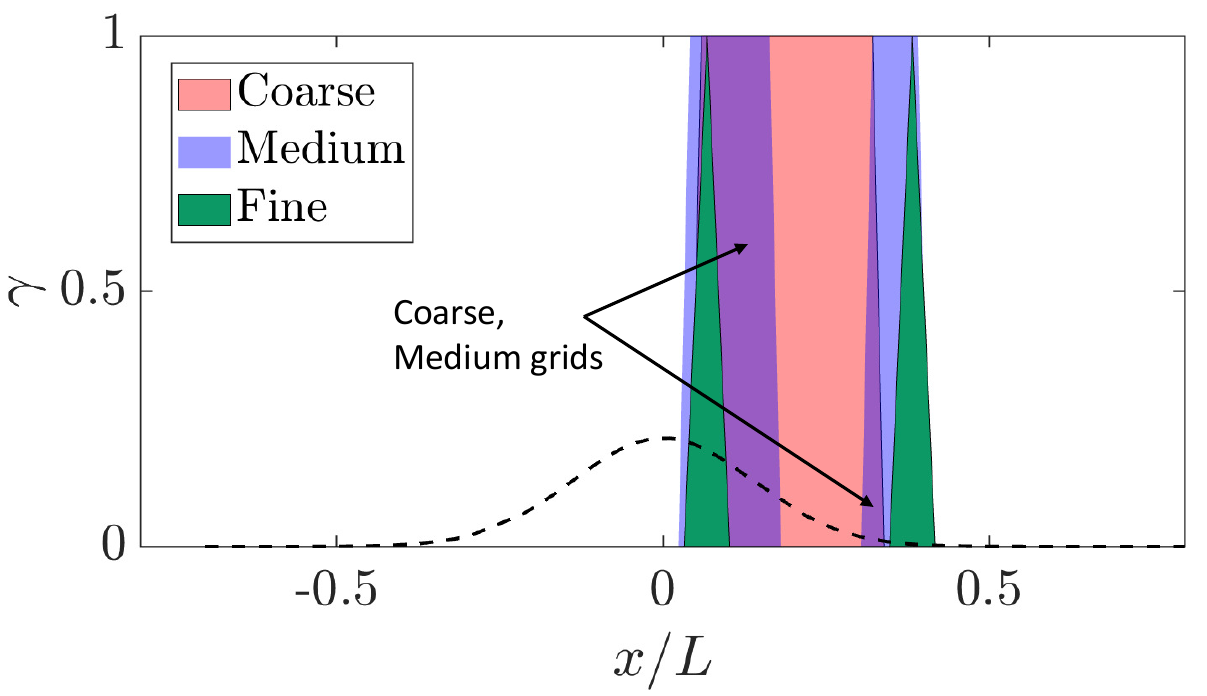}
    \caption{  The streamwise distribution of the averaged activity parameter $\gamma$ for the Boeing speed bump flow at $Re_L = 2 \times 10^6$. In the regions where $\gamma > 0 $, the sensor is active. In the experiments, the extent of the separation bubble is confined to within $x/L = 0.1 \; \mathrm{and} \; 0.4$. The black-dashed curve represents a schematic of the bump geometry. Note that the violet color represents the regions where the sensor is active on both the coarse and the medium grids.    }
    \label{fig:sensoraction}
\end{figure}

\noindent
Since the proposed wall model is aimed to capture non-equilibrium in the vicinity of a separation point, we expect that the proposed sensor only turns on in near separation regions. This is established in an \emph{a-posteriori} sense by plotting the activity, denoted by $\gamma =  \frac{1 + sgn ( \langle \chi \rangle )   }{2}$,
where $\langle \cdots \rangle $ represents the time and span averaging operator,  $\chi$ is the left side of the inequality in Equations \ref{eqn:historymodelterm4}-\ref{eqn:historymodelterm5} and $sgn(...)$ is the sign function. In effect, $\gamma$ represents the time-averaged spatial extent of the region of flow non-equilibrium as reported by the sensor. In the regions where the near-wall equilibrium assumption fails, the activity parameter $\gamma(x/L) = 0.0 $ and vice versa. Figure \ref{fig:sensoraction} presents the spatial distribution of $\gamma$ across the grid-refinement sequence. This plot confirms for this flow that the sensor augments the equilibrium wall model's prediction selectively in the regions of the strong adverse pressure gradient and inside the separation bubble. Further, as the grid is refined, the sensor reacts to the local flow state, and the activity is generally reduced compared to a coarser grid. On the ``fine grid", the sensor only becomes active at the mean separation point, and otherwise, the outer flow is resolved enough to largely be in equilibrium. 

\subsection{Reynolds number scaling of grid-point requirements of wall-modeled LES}


\begin{figure}[!ht]
    \centering
    \subfigure[]{
    \includegraphics[width=0.465\textwidth]{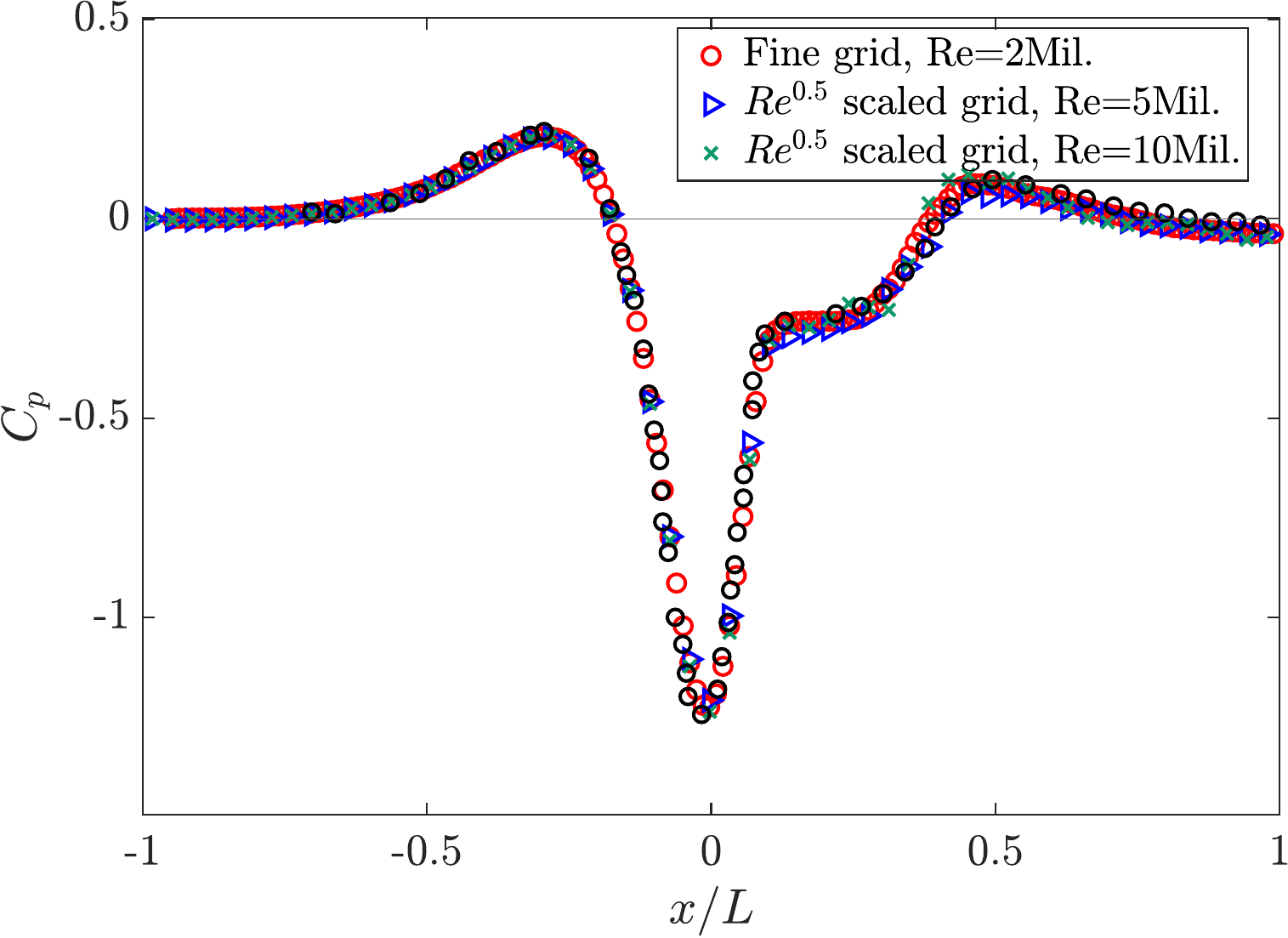}}
    \subfigure[]{
    \includegraphics[width=0.48\textwidth]{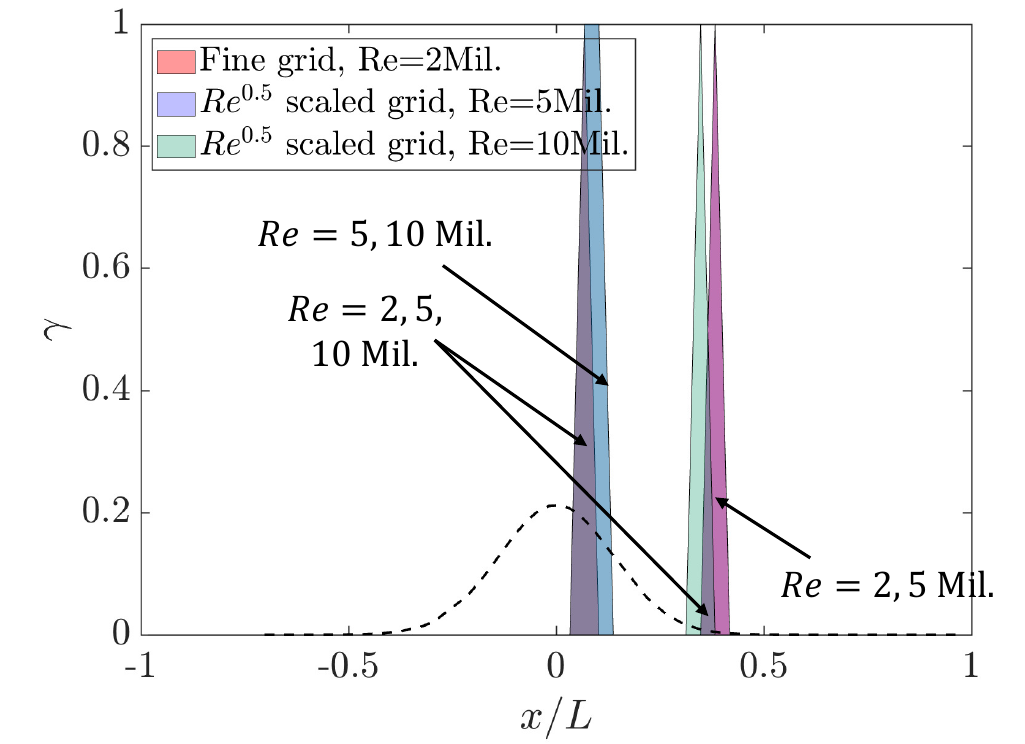}}
    \caption{  Streamwise distribution of the (a) surface pressure distribution coefficient, $C_p$, (b) activity parameter of the sensor at the wall for the flow over the spanwise-periodic Boeing speed bump as a function of the Reynolds number. In subfigure (a), the black dots represent the $C_p$ reported in the quasi-DNS of \citet{uzun2021high}. The fine grid at $Re_L = 2 \times 10^6$ is considered as the base grid in this plot. In subfigure (b), the black-dashed curve represents a schematic of the bump geometry, and the arrows point to the sensor becoming active for two or all three Reynolds numbers at the same $x/L$.  }
    \label{fig:cpRebump}
\end{figure}

\noindent
As noted before, the experiments of \citet{williams2020experimental} and  \citet{gray2022experimental,gray2022experimentalb} have both hypothesized an approximately Reynolds number independent extent of the separation bubble on the aft side of the bump at sufficiently high Reynolds numbers. Their experimental measurements were reported for $1.8 \times 10^6 \leq Re_L \leq 4 \times 10^6$. \citet{agrawal2023reynolds} performed a series of wall-modeled LES of the flow over the speed bump up to $Re_L = 10 \times 10^6$ and reported that the results converged towards the experimental $C_p$ trace (at $Re_L = 2 \times 10^6$) in the domain. These synthetic experiments are repeated to assess the grid-resolution scaling (with Reynolds number) of \emph{a-posteriori} wall-modeled LES employing the proposed wall model.\\

\noindent
The fine grid for the $Re_L = 2 \times 10^6$ flow is considered as the baseline grid. Per the scaling hypothesized in Section III(C), the grids are refined by a factor of $Re^{1/2}$ as the Reynolds number increases. 
Figure \ref{fig:cpRebump}(a) suggests that the predicted $C_p$ also compare well with each other and with the quasi-DNS \citep{uzun2021high} at $Re_L = 2 \times 10^6$. Figure \ref{fig:cpRebump}(b) confirms that the sensor reacts to the change in the Reynolds number and a larger region of flow non-equilibrium is highlighted as the Reynolds number is increased. These results suggest that at least for this flow, the \emph{a-priori} Reynolds number scaling of the grid resolution for capturing the separation bubble is also realized \emph{a-posteriori} with the proposed wall model.

\section{Wall Modeled LES of Bachalo-Johnson Axisymmetric Transonic Bump}

\noindent
Bachalo and Johnson \cite{bachalo1986transonic} conducted an experimental investigation of a canonical flow that produces a transonic shock-induced flow separation at a high chord-based Reynolds number ($Re_c = 2.76 \times 10^6$), along with the adverse-pressure gradient-induced smooth-body separation; both of which occur on the suction surface of transonic airfoils. In this experiment, an axisymmetric solid model with a spherical bump is placed in a flow with a transonic free-stream Mach number. The turbulent boundary layer that develops over the model's surface upstream of the bump interacts with the local supersonic flow over the bump, leading to the detachment of the boundary layer and flow separation. The experiments also included results at slightly different Mach numbers ($Ma = 0.85, \; 0.875, \; 0.90$), which produced significantly different flow separation and recovery regions. This finding has been since corroborated in a recent set of experiments at the Sandia National Labs \citep{lynch2023experimental}, albeit at a lower Reynolds number ($Re_c=1 \times 10^6$). We highlight that although the bump width-based Reynolds number of the Bachalo-Johnson flow is comparable to that over the Boeing speed bump, the incoming reference momentum (at $x/c=-0.25$) thickness-based Reynolds number for this flow is $Re_{\theta} \approx 12000$, which is an order of magnitude larger than the flow over the Boeing speed bump ($Re_{\theta} \approx 1400$). \\ 

\noindent
Previous efforts to simulate the flow over the Bachalo Johnson bump have generally been limited to the $Ma_e=0.875$ case. \citet{jespersen2016overflow} performed RANS computations using the OVERFLOW solver that did not predict flow separation. Later, \citet{spalart2017large} performed scale-resolving detached eddy simulations using approximately $\approx 400$ million control volumes. However, their solutions did not predict the flow separation either. More recently, \citet{lv2021discontinuous} have accurately predicted the surface-pressure distribution by performing wall-modeled LES using a discontinuous Galerkin framework. \citet{horstman1984prediction} have demonstrated the inability of RANS models to capture the sensitivity of the reattachment point to small variations in the Mach number. In this work, the sensitivity of this model to slight changes in the freestream conditions and the corresponding separation extent is examined by simulating the three slightly different Mach number flows. To the authors' knowledge, the ability of wall-modeled LES to respond to changes in Mach number in this flow has not been reported previously. \\

\subsection{Case setup and boundary conditions}

\begin{figure}[!ht]
    \centering
    \includegraphics[width=1.0\textwidth]{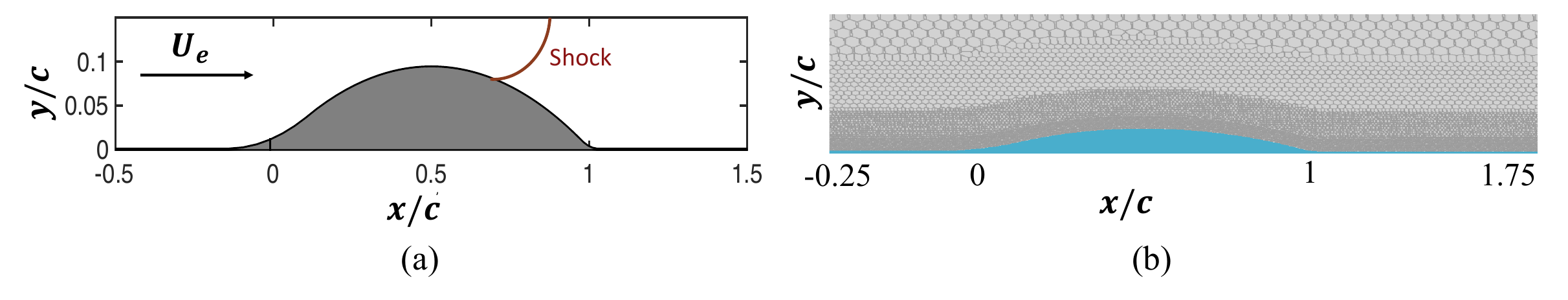}
    \caption{ A schematic of the (a) case setup and the (b) grid distribution along the streamwise direction for the transonic flow over the axisymmetric Bachalo-Johnson bump. It is highlighted that the grid is maintained to be agnostic to the location of the shock in the streamwise direction. Near-wall refinement is performed in layers of isotropically growing cells. The shock location for $Ma_e = 0.875$ flow is approximately at $x/c = 0.67$.   }
    \label{fig:geombachalo}
\end{figure}

\noindent
The Reynolds number of the flow is based on the freestream velocity, $U_e$, and the bump chord ($c$) and is equal to $Re_c = 2.76  \times 10^6$. The three different Mach numbers are $Ma_e=0.85, \; 0.875, \; 0.90$. A plug profile is fed at the inlet (located at $x/c = -4$), and the boundary layer is allowed to develop into a statistically equilibrated zero-pressure pressure gradient flow. The flow then encounters the bump at $x/c=0$ followed by a strong favorable pressure gradient region up to $x/c=0.5$. Thereafter, the flow experiences an adverse pressure gradient before it encounters the shock ($x/c \approx 0.67$) that leads to the flow separation downstream. The outlet is located at $x/c=10$ to allow the flow to recover from the separation. The simulation domain is a $60^\circ$ sector of the axisymmetric geometry, with the characteristic, non-reflective boundary conditions at the outlet. Figure \ref{fig:geombachalo}(a) shows the flow domain schematic in the near-bump region. Similarly, Figure \ref{fig:geombachalo}(b) shows the grid distribution over the bump surface; an isotropic, layered, near-wall refinement strategy is pursued to capture the smaller scales of the flow near the wall. Finally, the grid is also maintained to be agnostic to the streamwise location of the shock to challenge the wall-modeled LES methodology. The details of the grid resolution are provided in Table \ref{table:resbj}; the last column shows that in terms of the viscous length scale imposed by the pressure gradient ($l_p$), this flow is under-resolved.

\begin{table}
\centering
\begin{tabular}{ p{1.5cm}p{1.5cm}p{1.5cm}p{1.5cm}p{1.6cm}}

Mesh & $N_{cv}$ & max $\Delta / c$  & min $\Delta / c$ & max. $\Delta / l_p $  \\
\hline\noalign{\vspace{3pt}}
WMLES  & $31 $ Mil. & $2.5 \times 10^{-1} $ & $ 2 \times 10^{-3}$  & $\approx 70$   \\

\end{tabular}
\caption{Mesh parameter for a $60^{\circ}$ sector case of the axisymmetric, transonic flow over the bump in the experiments conducted by \citet{bachalo1986transonic}. Compared to the experimental boundary layer thickness in \citet{bachalo1986transonic}, at this grid resolution, the incoming boundary layer ($x/c \approx -0.25$) contained approximately $15$ grid points. In the last column, $l_p$ denotes the viscous length scale governed by the pressure gradient term, $l_p = \nu/u_p$.  }
\label{table:resbj}
\end{table}

\subsection{ Predictions with the proposed wall-model }
\label{sec:bjprediction}
\begin{figure}[!ht]
    \centering
    \includegraphics[width=0.8\textwidth]{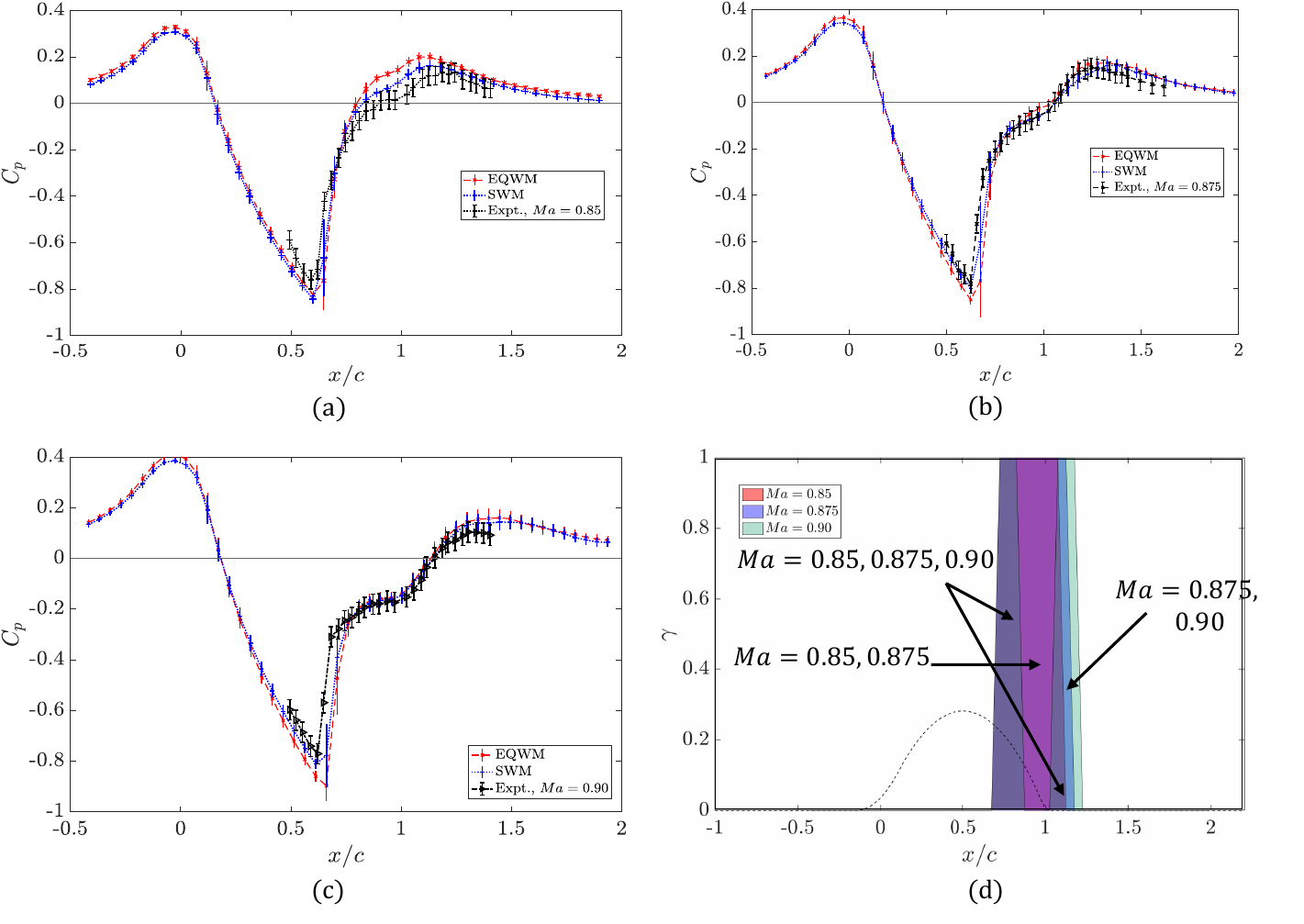}
    \caption{  A comparison of the predicted surface pressure coefficient, $C_p$, between the equilibrium wall model and the proposed model for the three Mach number cases, (a) $Ma_e=0.85$, (b) $Ma_e=0.875$ and (c) $Ma_e=0.90$. For reference, the $C_p$ traces from the experiments of \citet{bachalo1986transonic} are also reported. Subfigure (d) presents the activity parameter of the sensor at the wall,  the black-dashed curve is a schematic of the transonic bump geometry, and the arrows point to the sensor becoming active for two or all three Mach numbers at the same $x/c$.  }
    \label{fig:cpbachalo}
\end{figure}

\noindent
In the original Bachalo-Johnson experiment, the streamwise trace of $C_p$ was reported for all three Mach numbers, $Ma_e = 0.85, \; 0.875, \mathrm{and} \; 0.90$. In the current simulations, regardless of the varying Mach number, the same mesh, subgrid-scale, and wall model is employed. Figure \ref{fig:cpbachalo}(a)-(c) compares the prediction of the surface pressure resulting from the equilibrium wall model and the proposed model at $Ma = 0.85, \; 0.875, \mathrm{and} \; 0.90$ respectively. At the lowest Mach number ($Ma_e=0.85$ in subfigure (a)), the shock is not strong enough to separate the flow, and the adverse pressure gradient downstream of the shock leads to a flow separation (constant $C_p$ plateau) at $x/c \approx 0.75$. The proposed model, although slightly under-predicts the separation, is in more agreement with the experiments \citep{bachalo1986transonic} than the equilibrium model. The predictions on a more refined grid are found to be more favorable and are presented in Appendix III. The shock is strong enough to lead to flow separation at its foot at the two higher Mach numbers in subfigures (b)-(c). Both the models reasonably predict this separation and the subsequent reattachment location (although at $Ma_e=0.875$, the proposed model is slightly better than the equilibrium wall model). \\

\noindent
The size of the separation bubble also increases with the Mach number, possibly due to an increased effect of the pressure gradient (the Von Karman integral equation suggests that the skin friction is dependent on the product of an explicit function of the pressure gradient, $f(dP/dx)$ and the Mach number, $Ma^2_e$). It was verified that the viscous resolution of the matching location for our simulations is $\mathcal{O}(100)$ in the upstream, zero-pressure gradient boundary layer region ($x/c \leq 0$) and that simulations using a no-slip boundary condition produce inaccurate solutions. Similar to Figure \ref{fig:sensoraction}, the activity of the proposed sensor is plotted in Figure \ref{fig:cpbachalo}(d). The sensor becomes active only at or after the foot of the shock ($x/c \approx 0.67$) for all three Mach numbers, which is in line with the location of the onset of flow separation. Additionally, the sensor responds to a change in the Mach number, remaining active for a slightly larger $x/c$ location as the Mach number increases. This is consistent with the fact that the reattachment point also moves downstream with Mach number. \\



\begin{figure}[!ht]
    \centering
    \subfigure[]{
    \includegraphics[width=0.3\textwidth]{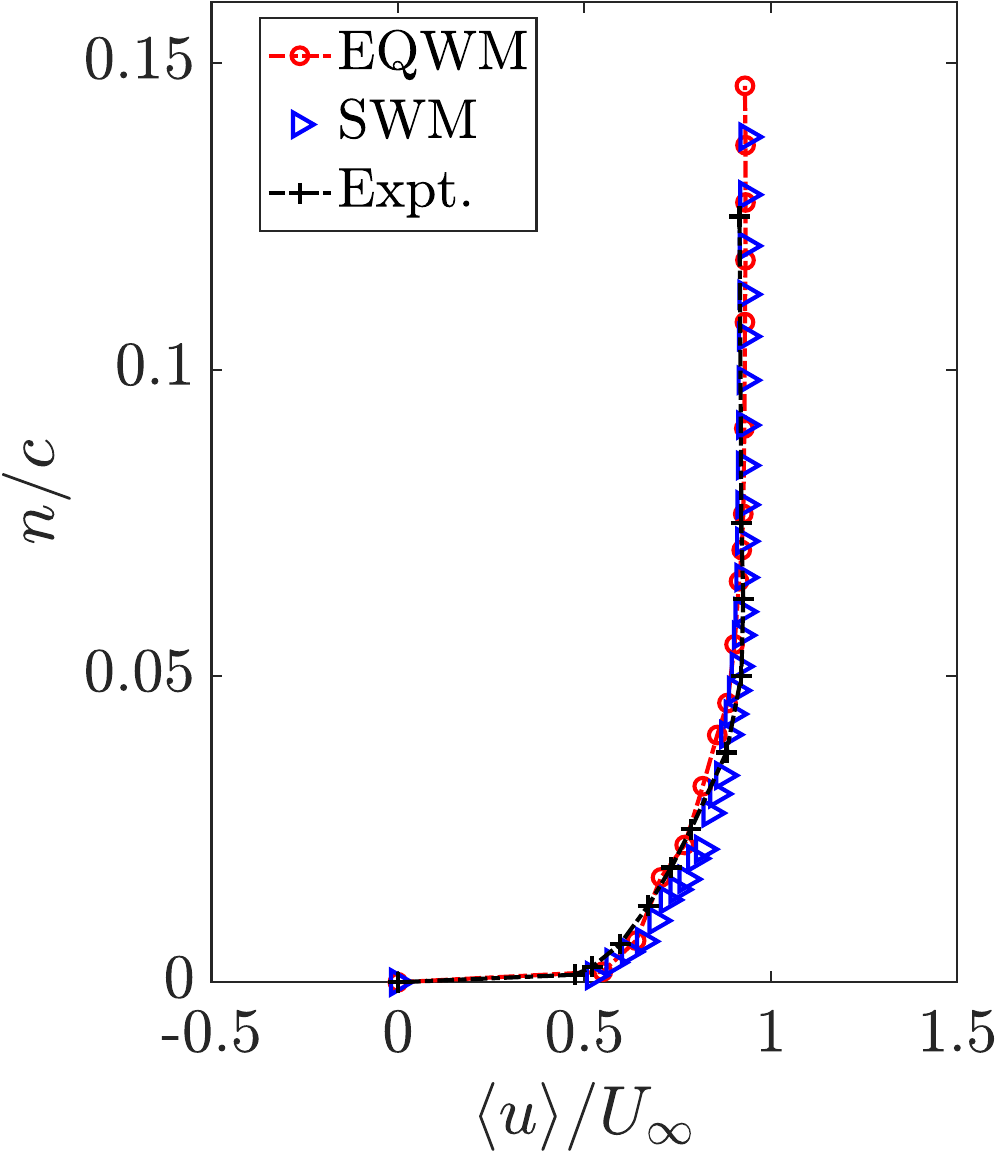} }
    \subfigure[]{
    \includegraphics[width=0.3\textwidth]{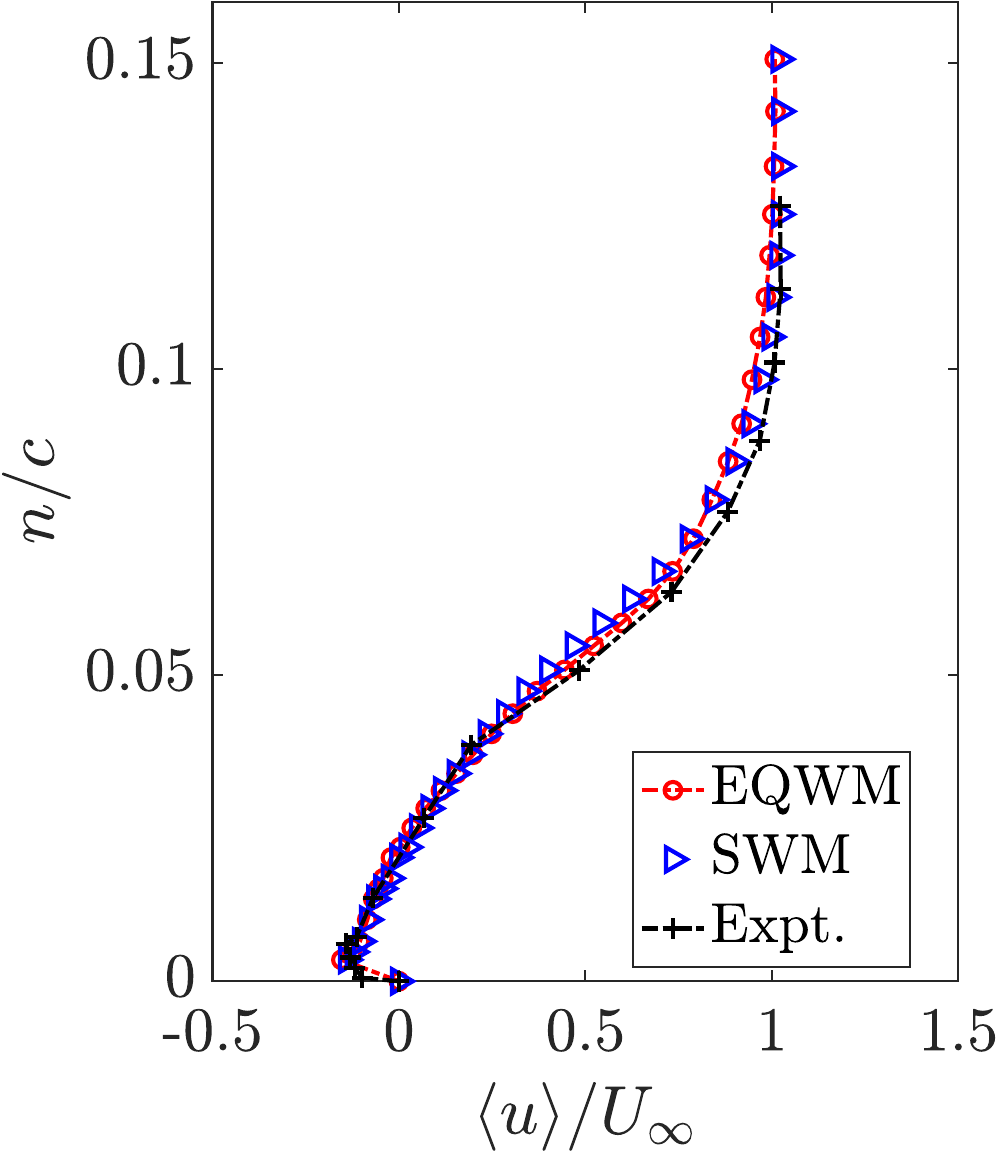} }
    \subfigure[]{
    \includegraphics[width=0.3\textwidth]{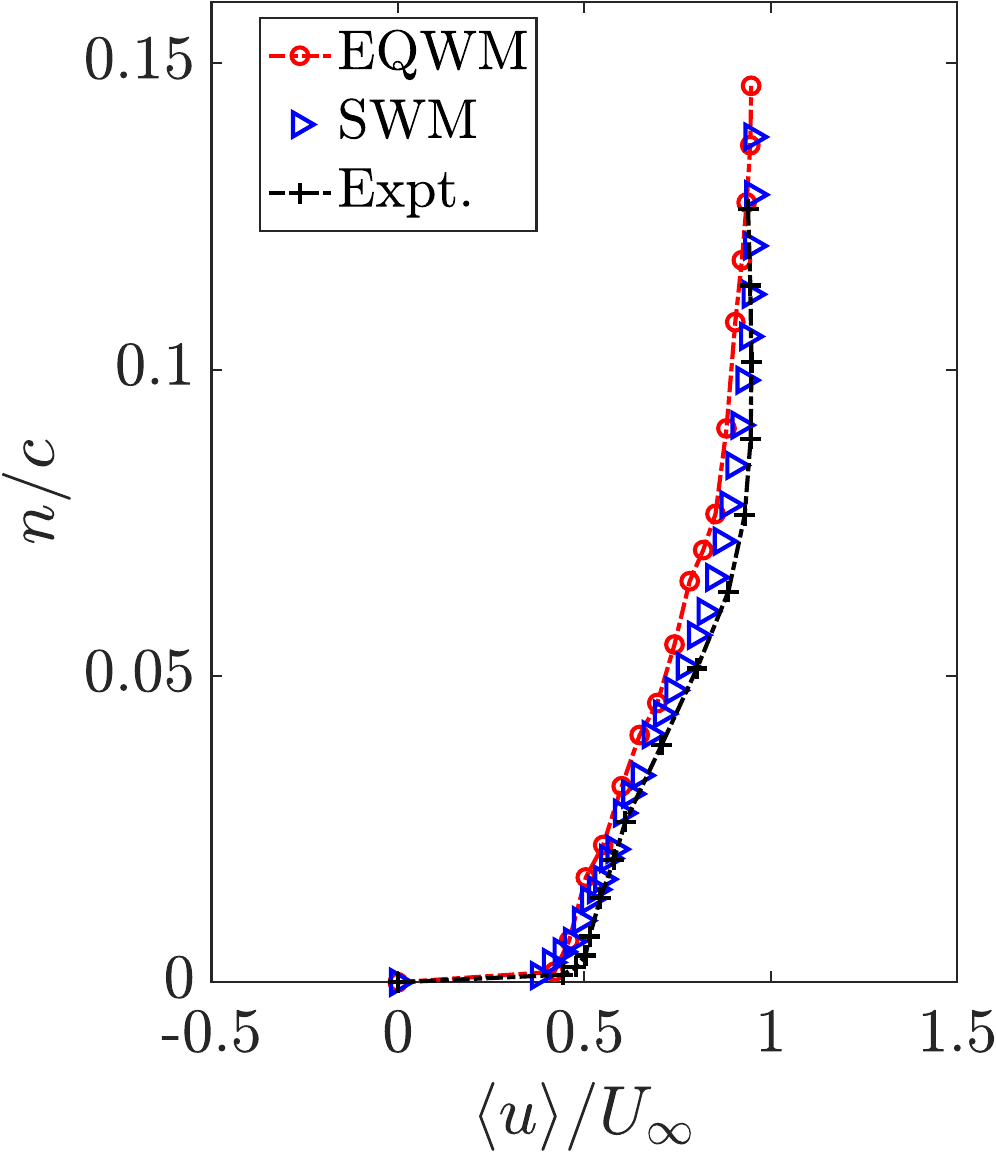} }
    \caption{Wall normal (along the direction orthogonal to the axis of the bump) profiles of the normalized streamwise velocity, $\langle u \rangle / U_{\infty} $, at (a) $x/c=-0.25$, (b) $x/c=1.00$, and (c) $x/c=1.375$ for flow over the transonic Bachalo-Johnson bump at $Re_c=2.76 \times 10^6$ and $Ma_e=0.875$. The reference experiments shown in black are from the experiments of \citet{bachalo1986transonic}. The three selected locations represent the upstream region before the bump, the separated region at the foot of the bump, and the downstream region past the reattachment point, respectively.  }
    \label{fig:bachalovels}
\end{figure}

\noindent
The experiments of \citet{bachalo1986transonic} provide wall-normal profiles of the streamwise-velocity at several $x/c$ stations for the $Ma_e=0.875 $ case. The prediction of the outer flow from wall-modeled LES is assessed using these velocity profiles. Figure \ref{fig:bachalovels} compares these predictions between the two models at three stations, $x/c=-0.25, \; 1.00, \; 1.375 $, which are representative of the upstream, separated, and recovery regions, respectively. At $x/c=-0.25$, both models produce similarly accurate velocity profiles, which is expected as the flow is in ``equilibrium". In the separated region ($x/c=1.00$) as well, the predicted velocity profiles from both models compare well with the experiment. These profiles also reasonably predict the two inflection points, which lead to the formation of embedded shear layers (see \citet{schatzman2017experimental}) in separated flows. In the post-reattachment and recovery region ($x/c=1.375$), the proposed model slightly improves the prediction of the mean velocity in comparison to the equilibrium wall model. Overall, it is concluded that the predictions of the mean velocities from wall-modeled LES are in good agreement with the reference experimental data \citep{bachalo1986transonic}.  \\

\subsection{Dependence of flow separation on Mach number }
\begin{figure}[!ht]
    \centering
    \includegraphics[width=0.8\textwidth]{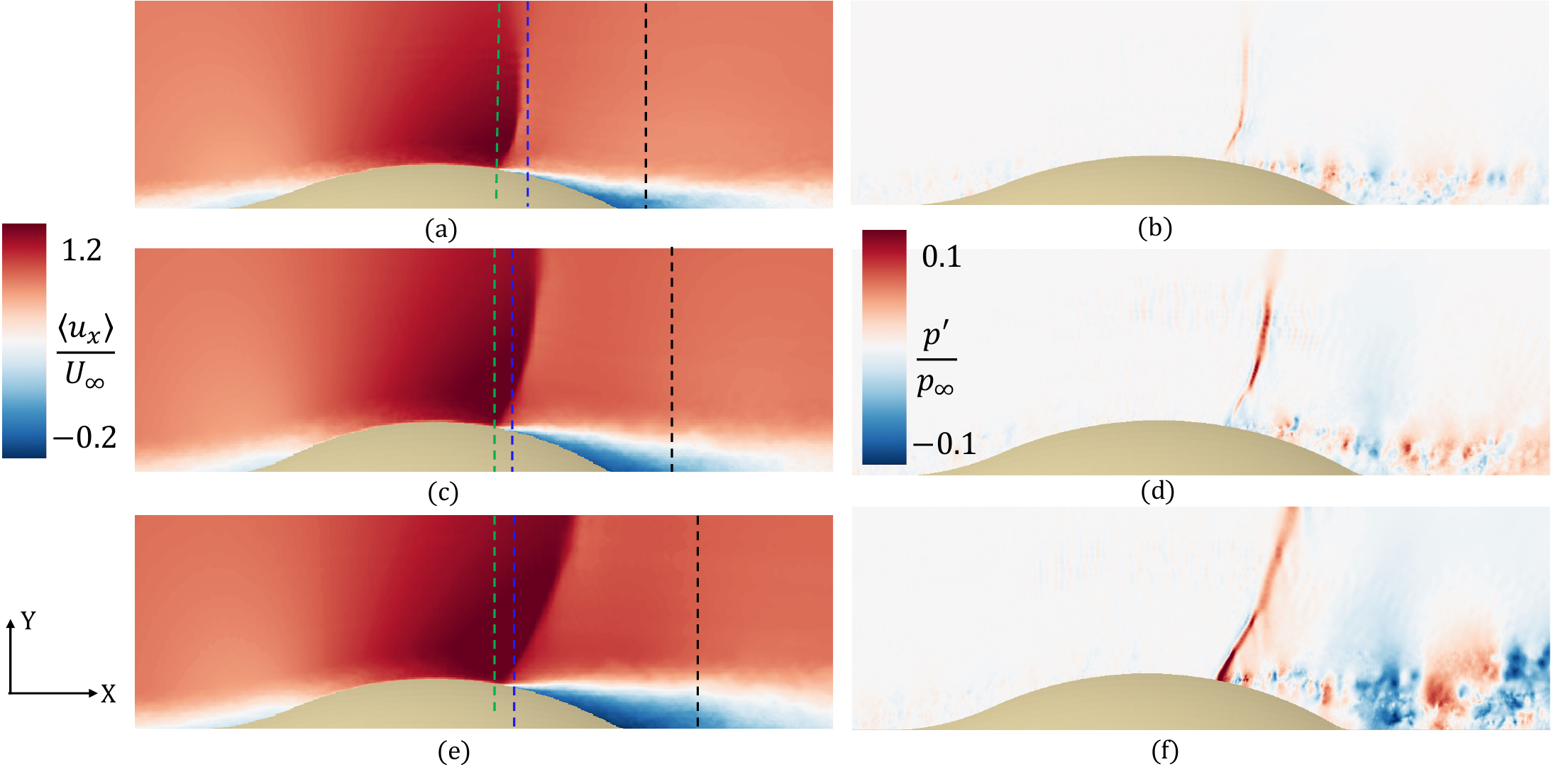}
    \caption{  Contours of time-averaged mean velocity (denoted as $\langle u_x \rangle $) at (a) $Ma_e=0.85$, (c) $Ma_e=0.875$ and (e) $Ma_e=0.90$ from wall modeled LES using the proposed wall model. The green, purple, and black dashed lines correspond to the location of the shock, the point of separation, and the point of reattachment, respectively. Subfigures (b), (d), and (f) present contours of the normalized, instantaneous pressure fluctuations ($p^{\prime}/p_{\infty}$) for flows at $Ma_e=0.85$, $Ma_e=0.875$ and $Ma_e=0.90$ respectively.   }
    \label{fig:vizbachalo}
\end{figure}

\noindent 
Figures \ref{fig:vizbachalo}(a), (c), and (e) present contours of the time-averaged streamwise velocity for the three Mach numbers at $0^{\circ}$ azimuthal angle. The green, purple, and black dashed lines correspond to the location of the shock, the point of separation, and the point of reattachment, respectively. The shock location is found to be relatively insensitive to the Mach number; however, the shock becomes more oblique, as expected, with the Mach number. The separation point for $Ma_e=0.85$ is slightly downstream of the two higher $Ma_e$ cases, which separate approximately at the foot of the shock. The separated shear layer (marked by the white contours beyond the separation point) is more lifted, and then attaches with the bump wall further downstream as the Mach number increases. Figures \ref{fig:vizbachalo} (b), (d), and (f) present the contours of the instantaneous pressure fluctuations for the flow cases at $Ma_e=0.85$, $Ma_e=0.875$ and $Ma_e=0.90$ respectively. The increasing obliqueness of the shock with the Mach number is also visible in these figures. The separated region downstream of the shock produces larger (in magnitude) pressure fluctuations as the Mach number increases. This is due to the unsteady shedding of the shear layer, which is correlated to the formation of a larger separation bubble. \\

\section{Concluding Remarks}

\noindent
In this work, a new wall model is proposed specifically aimed at identifying regions of strong near-wall flow non-equilibrium (in an averaged sense) that are inaccurately modeled using equilibrium boundary layer approximations. This region is presently ascertained in regions where the viscous pressure gradient velocity ($u_p$) is significant compared to the friction velocity ($u_{\tau}$).  In these regions, the thin boundary layer equations used to approximate the wall stress are solved using the approximation of \citet{kamogawa2023ordinary} for 
the combined pressure gradient/convective terms. The performance of the proposed wall model is assessed on the flow over the Boeing Speed bump as a canonical case exhibiting smooth-body separation. Improvements in the grid-point scaling of the proposed model with Reynolds number compared to the equilibrium wall model are both hypothesized and then demonstrated in an \emph{a posteriori} sense. The responsiveness of the model to adjust to small changes in freestream Mach number has been examined by performing wall-modeled LES of the transonic flow over the axisymmetric Bachalo-Johnson bump with the proposed model capturing the large differences in the $C_p$ in the separation region (maximum difference between the extreme $Ma_e$ cases being $\Delta C_p \approx 0.25$) due to small changes in the Mach number (6\% change). Overall, the results in this article demonstrate an improvement in the quantities of interest (such as the skin friction, surface pressure, and the mean velocity profiles) in boundary layers experiencing pressure gradients that lead to a smooth-body separation. 
  

\section*{Acknowledgements}

This work was supported by NASA's Transformational Tools and Technologies project under grant number 80NSSC20M0201 and by Boeing Research \& Technology. This research used resources from the Oak Ridge Leadership Computing Facility, which is a DOE Office of Science User Facility supported under Contract DE-AC05-00OR22725. 

\small
\section*{Appendix I: Connection between the velocity-scale ratio, $\frac{u_p}{u_{\tau}}$, and the critical, viscously stable inner layer \label{sec:appendix1}
}
\noindent
\citet{nickels2004inner} and \citet{knopp2021experimental} argued that the relevant velocity scale in the near-wall region for a boundary layer experiencing pressure gradients is the local shear stress instead of the wall shear stress. \citet{lozano2019characteristic} also corroborated this scaling for a turbulent channel flow with prescribed mean velocity profiles. \citet{nickels2004inner} suggested that the critical value of the distance from the wall, $y^{+}_c$ where the flow becomes viscously ``unstable" and the effect of the pressure gradient becomes relevant is given by the relation, 
\begin{equation} 
  \frac{ |\vec{u_p}| ^3 }{|\vec{u_{\tau}}| ^3}  \; y_c^{+3} + y_c^{+2} - R_c^2  = 0 
\label{eqn:instability}
\end{equation}
where $R_c \approx 12$ is the universal critical Reynolds number that determines the edge of the stable-viscous layer. For a zero pressure gradient boundary layer, $y^{+}_c \sim 12$, coincides with the location of the peak production \cite{kim1987turbulence}. The height of the stable viscous layer will be significantly affected by the presence of pressure gradients if ${y^+_c} \frac{ |\vec{u_p}| ^3 }{|\vec{u_{\tau}}| ^3} = {y^{p}_c} \frac{ |\vec{u_p}| ^2 }{|\vec{u_{\tau}}| ^2}  \sim  1$ or $  \frac{ |\vec{u_p}| }{|\vec{u_{\tau}}| } \sim  \sqrt{\frac{1}{y^p_c}} \sim \mathcal{O}(1) $. 

\section*{Appendix II: Sensitivity of prediction of flow separation to sensor-activity parameter }
\label{sec:appendix2}

\begin{figure}[!ht]
    \centering
    \includegraphics[width=0.4\textwidth]{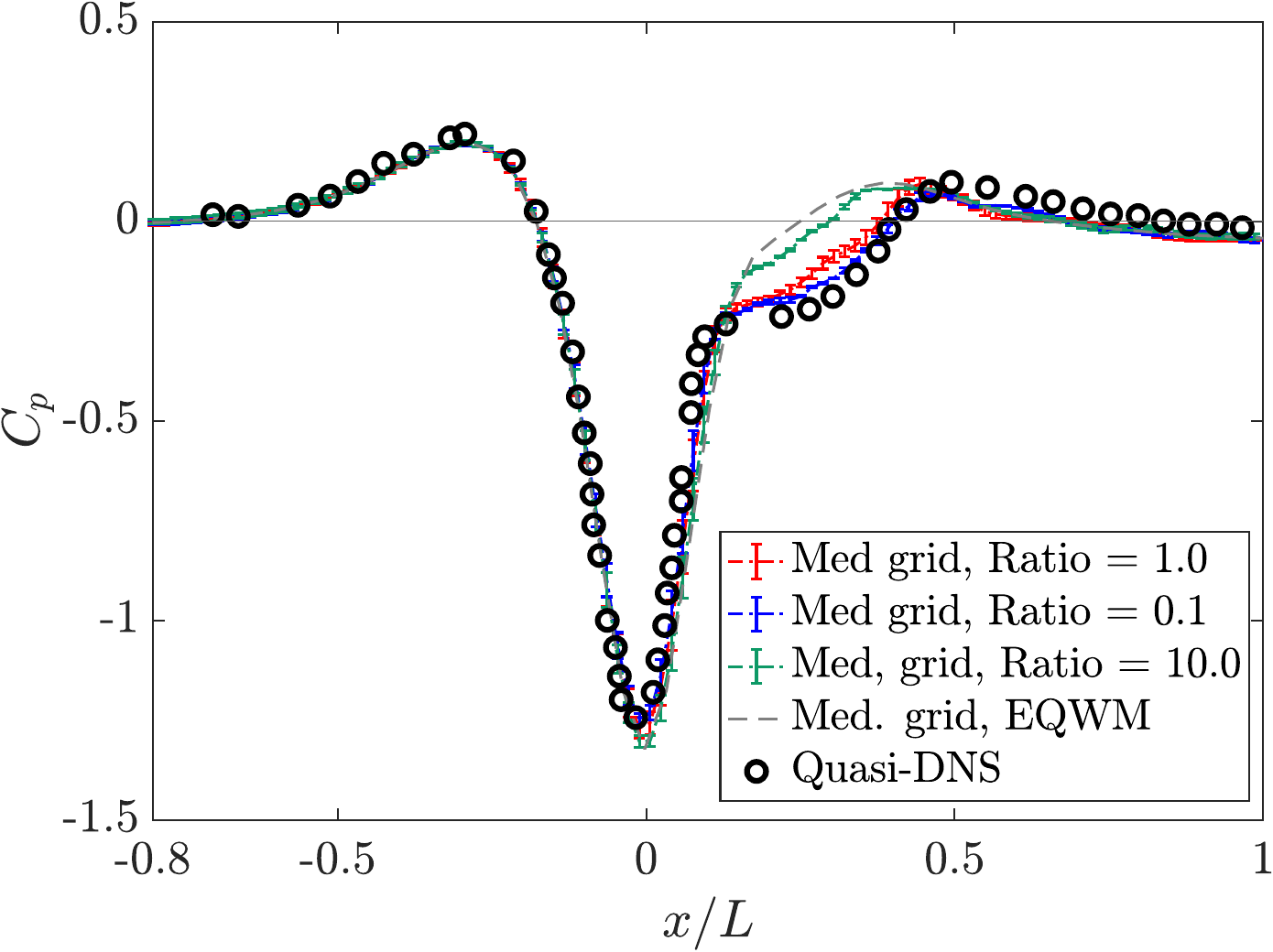}
    \caption{ Sensitivity of the streamwise surface pressure coefficient, $C_p$ to the choice of the sensor activity parameter (ratio of $|u_p|/|u_\tau | $). The flow considered here is that over the Boeing speed bump at $Re_L = 2 \times 10^6$. The reference quasi-DNS data is from \citet{uzun2021high}.  }
    \label{fig:stbsenstivityquestion}
\end{figure}

In Equation \ref{eqn:historymodelterm3}, the ratio of  $ \frac{ |\vec{u^2_p}|}{|\vec{u^2_{\tau}}|} $ was bounded by the upper limit of $\frac{\mathcal{O}(1)}{ y^p_1 [-A log(y_1^p) +B]  }$ for  $ y^p_1 \geq \mathcal{O}(10)$, resulting in $ \frac{ |\vec{u^2_p}|}{|\vec{u^2_{\tau}}|}  \sim \mathcal{O}(1)$. In this appendix, we test the sensitivity of the \emph{a-posteriori} prediction of wall-modeled LES to the choice of this lower bound. Specifically, we consider the medium grid on the flow over the Boeing speed bump at $Re_L = 2 \times 10^6$ (see Table \ref{table:resbump2}). Figure \ref{fig:stbsenstivityquestion} shows that upon decreasing the threshold parameter ratio by an order of magnitude, the prediction of the pressure trace improves; as the flow approaches reattachment, the prediction of $C_p$ is slightly better. Upon increasing the desired threshold ratio up to 10.0, the separation decreases significantly; the results are only slightly improved compared to the equilibrium wall model (which is the limit of the proposed model as the threshold ratio $ \frac{ |\vec{u^2_p}|}{|\vec{u^2_{\tau}}|} \rightarrow \infty $). We also highlight that a threshold ratio larger than one is unphysical since, in that case, $\mathcal{H}$ will exceed one. Secondly, in the vicinity of separation and reattachment points, this ratio is bound to reach infinitely large values eventually. 

\section*{Appendix III: Grid convergence for transonic Bachalo-Johnson bump flow }
\label{sec:appendix3}

\begin{table}
\centering
\begin{tabular}{ p{1.5cm}p{1.5cm}p{1.5cm}p{1.5cm}p{1.6cm} }

Mesh & $N_{cv}$ & max $\Delta / c$  & min $\Delta / c$ & max. $\Delta / l_p $  \\
\hline\noalign{\vspace{3pt}}
Coarse  & $31 $ Mil. & $2.5 \times 10^{-1} $ & $ 2 \times 10^{-3}$ & $\approx 70$ \\
Fine  & $120 $ Mil. & $2.5 \times 10^{-1} $ & $ 1 \times 10^{-3}$ & $\approx 35$ \\

\end{tabular}
\caption{Mesh parameters for a $60^{\circ}$ sector case of the transonic flow at $Ma_e=0.85$ over the bump in the experiments conducted by \citet{bachalo1986transonic}. In the last column, $l_p$ denotes the viscous length scale governed by the pressure gradient term, $l_p = \nu/u_p$. }
\label{table:resbj2}
\end{table}

\begin{figure}[!ht]
    \centering
    \subfigure[]{
    \includegraphics[width=0.43\textwidth]{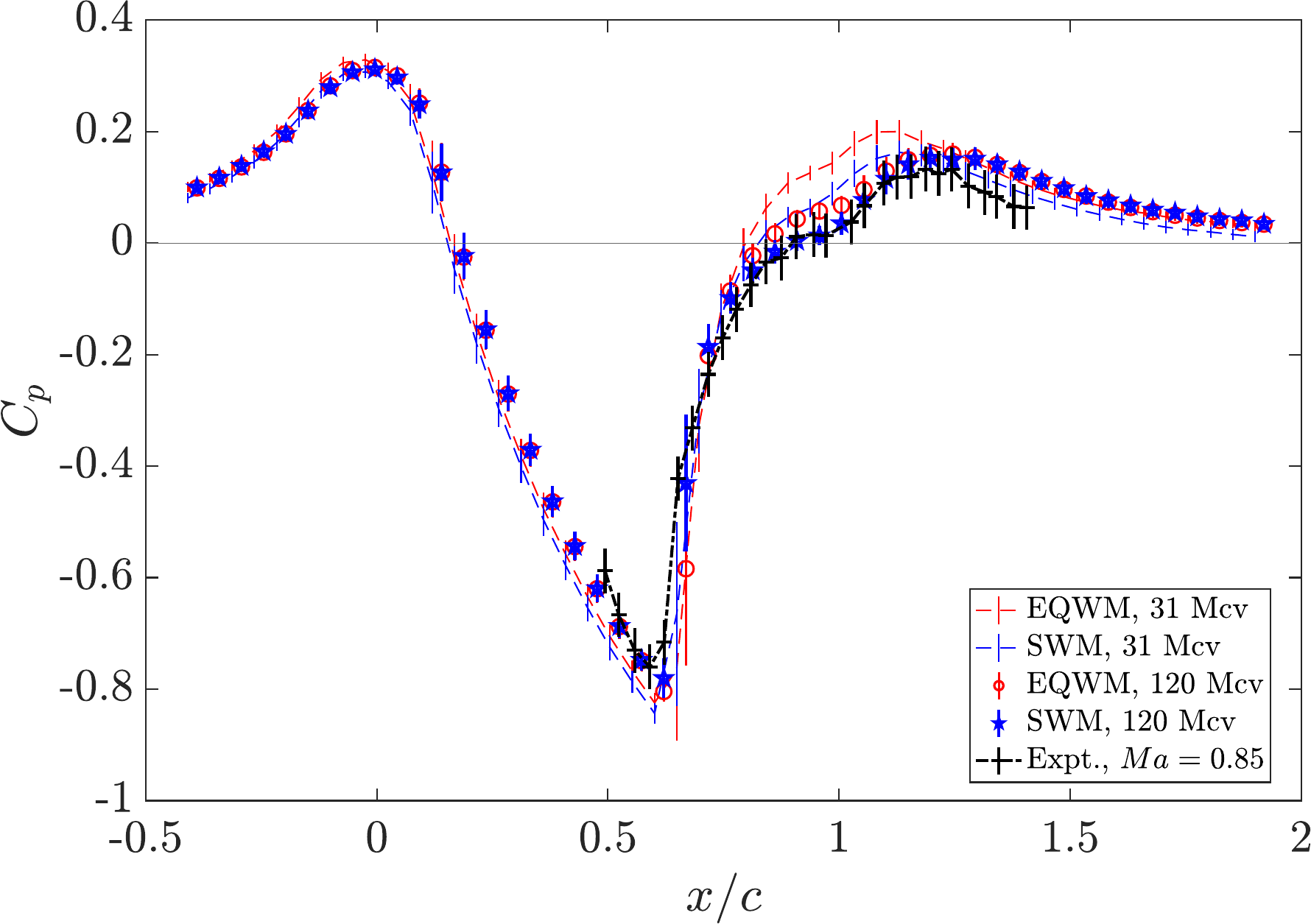}}
    \subfigure[]{
    \includegraphics[width=0.45\textwidth]{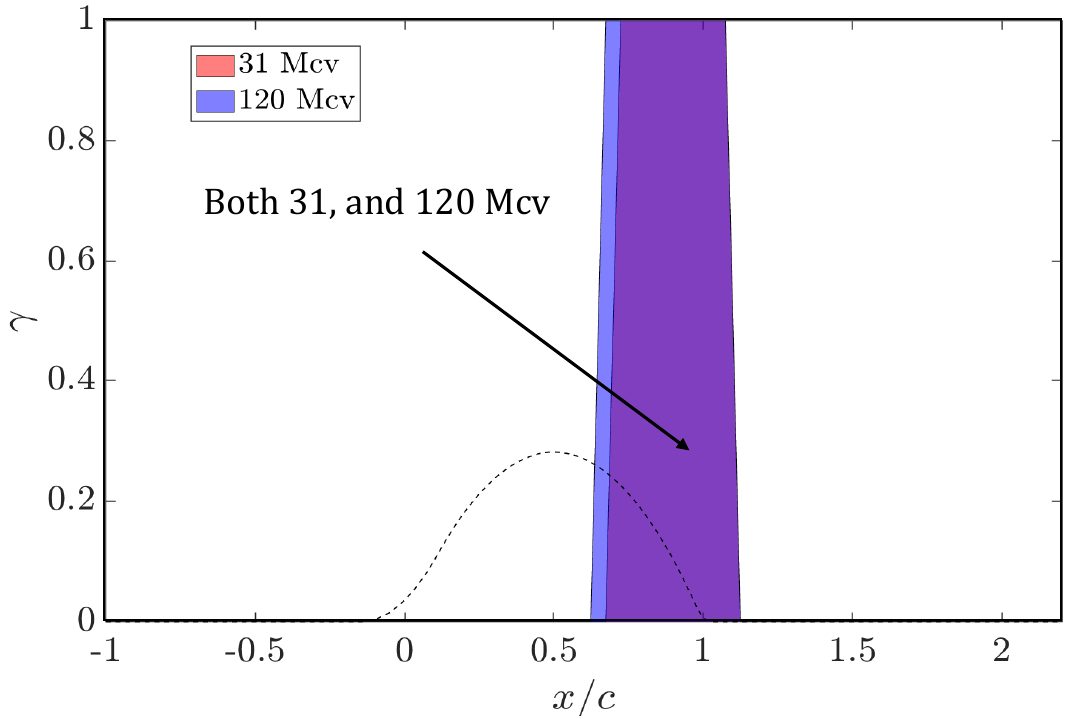}}
    \caption{  Contours of time-averaged mean velocity (in the X direction, or at $0^{\circ}$ azimuthal angle, denoted as $\langle u_x \rangle $) from wall modeled LES (using 31 million control volumes) leveraging (a) the equilibrium wall model and (b) the proposed wall model.} 
    \label{fig:cpbj85finer}
\end{figure}

As remarked in Section \ref{sec:bjprediction}, our wall-modeled LES results for the flow over the Bachalo-Johnson bump at the lowest Mach number ($Ma_e=0.85$) slightly unpredict the flattening of the $C_p$ curve in the separated region. In Figure \ref{fig:cpbj85finer}(a), predictions from a more refined grid (2x refined homothetically compared to the 31 Mcv grid in Seciton \ref{sec:bjprediction}; refer to Table \ref{table:resbj2} for more details) are presented for both the equilibrium wall model and the proposed wall model. It is apparent that the proposed model improves the prediction of $C_p$, especially on the coarser grid. The prediction of the two models on the finer grid is more favorable with the experiments \citep{bachalo1986transonic}. In Figure \ref{fig:cpbj85finer}(b), the activity of the sensor is plotted across the two grids. On the finer grid, the sensor switches on slightly upstream in comparison to the coarser grid as the separation moves slightly upstream. \\

\small

\begin{thebibliography}{69}
\providecommand{\natexlab}[1]{#1}
\providecommand{\url}[1]{\texttt{#1}}
\expandafter\ifx\csname urlstyle\endcsname\relax
  \providecommand{\doi}[1]{doi: #1}\else
  \providecommand{\doi}{doi: \begingroup \urlstyle{rm}\Url}\fi

\bibitem[Pope(2000)]{pope2000turbulent}
Stephen Pope.
\newblock \emph{Turbulent flows}.
\newblock Cambridge university press, 2000.

\bibitem[Smagorinsky(1963)]{smagorinsky1963general}
Joseph Smagorinsky.
\newblock General circulation experiments with the primitive equations: I. the
  basic experiment.
\newblock \emph{Monthly Weather Review}, 91\penalty0 (3):\penalty0 99--164,
  1963.

\bibitem[Germano et~al.(1991)Germano, Piomelli, Moin, and
  Cabot]{germano1991dynamic}
Massimo Germano, Ugo Piomelli, Parviz Moin, and William~H. Cabot.
\newblock A dynamic subgrid-scale eddy viscosity model.
\newblock \emph{Physics of Fluids A: Fluid Dynamics}, 3:\penalty0 1760--1765,
  1991.

\bibitem[Nicoud et~al.(2011)Nicoud, Toda, Cabrit, Bose, and Lee]{nicoud_sigma}
Franck Nicoud, Hubert~Baya Toda, Olivier Cabrit, Sanjeeb Bose, and Jungil Lee.
\newblock Using singular values to build a subgrid-scale model for large eddy
  simulations.
\newblock \emph{Physics of Fluids}, 23\penalty0 (8):\penalty0 085106, 2011.

\bibitem[Vreman et~al.(1994)Vreman, Geurts, and Kuerten]{vreman1994formulation}
Bert Vreman, Bernard Geurts, and Hans Kuerten.
\newblock On the formulation of the dynamic mixed subgrid-scale model.
\newblock \emph{Physics of Fluids}, 6\penalty0 (12):\penalty0 4057--4059, 1994.

\bibitem[Rozema et~al.(2015)Rozema, Bae, Moin, and
  Verstappen]{rozema2015minimum}
Wybe Rozema, Hyun~J Bae, Parviz Moin, and Roel Verstappen.
\newblock Minimum-dissipation models for large-eddy simulation.
\newblock \emph{Physics of Fluids}, 27\penalty0 (8):\penalty0 085107, 2015.

\bibitem[Agrawal et~al.(2022-07)Agrawal, Whitmore, Griffin, Bose, and
  Moin]{agrawal2022non}
Rahul Agrawal, Michael~P. Whitmore, Kevin~P. Griffin, Sanjeeb~T. Bose, and
  Parviz Moin.
\newblock Non-{B}oussinesq subgrid-scale model with dynamic tensorial
  coefficients.
\newblock \emph{Physical Review Fluids}, 7:\penalty0 074602, 2022-07.

\bibitem[Chenyue et~al.(2021)Chenyue, Zelong, Jianchun,
  et~al.]{chenyue2021artificial}
Xie Chenyue, Yuan Zelong, Wang Jianchun, et~al.
\newblock Artificial neural network-based subgrid-scale models for large-eddy
  simulation of turbulence.
\newblock \emph{Chinese Journal of Theoretical and Applied Mechanics},
  53\penalty0 (1):\penalty0 1--16, 2021.

\bibitem[Zhou et~al.(2019)Zhou, He, Wang, and Jin]{zhou2019subgrid}
Zhideng Zhou, Guowei He, Shizhao Wang, and Guodong Jin.
\newblock Subgrid-scale model for large-eddy simulation of isotropic turbulent
  flows using an artificial neural network.
\newblock \emph{Computers \& Fluids}, 195:\penalty0 104319, 2019.

\bibitem[Bae et~al.(2019)Bae, Lozano-Dur\'{a}n, Bose, and Moin]{bae2019}
H.~Jane Bae, Adri\'{a}n Lozano-Dur\'{a}n, Sanjeeb~T. Bose, and Parviz Moin.
\newblock Dynamic slip wall model for large-eddy simulation.
\newblock \emph{Journal of Fluid Mechanics}, 859:\penalty0 400--432, 2019.
\newblock ISSN 14697645.

\bibitem[Yang et~al.(2015)Yang, Sadique, Mittal, and
  Meneveau]{yang2015integral}
XIA Yang, J~Sadique, R~Mittal, and Charles Meneveau.
\newblock Integral wall model for large eddy simulations of wall-bounded
  turbulent flows.
\newblock \emph{Physics of Fluids}, 27\penalty0 (2):\penalty0 025112, 2015.

\bibitem[Choi and Moin(2012)]{choi2012grid}
Haecheon Choi and Parviz Moin.
\newblock Grid-point requirements for large eddy simulation: Chapman$'s$
  estimates revisited.
\newblock \emph{Physics of Fluids}, 24\penalty0 (1):\penalty0 011702, 2012.

\bibitem[Deardorff(1970)]{deardorff1970numerical}
James~W. Deardorff.
\newblock A numerical study of three-dimensional turbulent channel flow at
  large {R}eynolds numbers.
\newblock \emph{Journal of Fluid Mechanics}, 41\penalty0 (2):\penalty0
  453--480, 1970.

\bibitem[Cabot and Moin(2000)]{cabot2000approximate}
William Cabot and Parviz Moin.
\newblock Approximate wall boundary conditions in the large-eddy simulation of
  high {R}eynolds number flow.
\newblock \emph{Flow, Turbulence and Combustion}, 63\penalty0 (1):\penalty0
  269--291, 2000.

\bibitem[Goc et~al.(2021)Goc, Lehmkuhl, Park, Bose, and Moin]{goc2021large}
Konrad~A. Goc, Oriol Lehmkuhl, George~Ilhwan Park, Sanjeeb~T. Bose, and Parviz
  Moin.
\newblock Large eddy simulation of aircraft at affordable cost: a milestone in
  computational fluid dynamics.
\newblock \emph{Flow}, 1:\penalty0 E14, 2021.

\bibitem[Goc et~al.(0)Goc, Moin, Bose, and Clark]{hlcrmkonrad}
Konrad~A. Goc, Parviz Moin, Sanjeeb~T. Bose, and Adam~M. Clark.
\newblock Wind tunnel and grid resolution effects in large-eddy simulations of
  the high-lift common research model.
\newblock \emph{Journal of Aircraft}, 0\penalty0 (0):\penalty0 1--13, 0.

\bibitem[Bose and Park(2018)]{bose2018wall}
Sanjeeb~T. Bose and George~Ilhwan Park.
\newblock Wall-modeled large-eddy simulation for complex turbulent flows.
\newblock \emph{Annuaul Review of Fluid Mechanics}, 50:\penalty0 535--561,
  2018.

\bibitem[Agrawal et~al.(2023{\natexlab{a}})Agrawal, Bose, and
  Moin]{agrawal2023reynolds}
Rahul Agrawal, Sanjeeb Bose, and Parviz Moin.
\newblock Reynolds number dependence of length scales governing turbulent flow
  separation with application to wall-modeled large-eddy simulations.
\newblock \emph{arXiv preprint arXiv:2401.00075}, 2023{\natexlab{a}}.

\bibitem[Wang and Moin(2002)]{wang2002dynamic}
Meng Wang and Parviz Moin.
\newblock Dynamic wall modeling for large-eddy simulation of complex turbulent
  flows.
\newblock \emph{Physics of Fluids}, 14:\penalty0 2043--2051, 2002.

\bibitem[Balaras et~al.(1996)Balaras, Benocci, and Piomelli]{balaras1996two}
Elias Balaras, Carlo Benocci, and Ugo Piomelli.
\newblock Two-layer approximate boundary conditions for large-eddy simulations.
\newblock \emph{AIAA {J}ournal}, 34\penalty0 (6):\penalty0 1111--1119, 1996.

\bibitem[Hickel et~al.(2012)Hickel, Touber, Bodart, and
  Larsson]{hickel2013parametrized}
S~Hickel, E~Touber, J~Bodart, and J~Larsson.
\newblock A parametrized non-equilibrium wall-model for large-eddy simulations.
\newblock In \emph{Proceedings of the Summer Program}, page 127. Citeseer,
  2012.

\bibitem[Park and Moin(2014)]{park2014improved}
George~Ilhwan Park and Parviz Moin.
\newblock An improved dynamic non-equilibrium wall-model for large eddy
  simulation.
\newblock \emph{Physics of Fluids}, 26\penalty0 (1):\penalty0 37--48, 2014.

\bibitem[Park and Moin(2016)]{park2016numerical}
George~Ilhwan Park and Parviz Moin.
\newblock Numerical aspects and implementation of a two-layer zonal wall model
  for les of compressible turbulent flows on unstructured meshes.
\newblock \emph{Journal of Computational Physics}, 305:\penalty0 589--603,
  2016.

\bibitem[Slotnick et~al.(2014)Slotnick, Khodadoust, Alonso, Darmofal, Gropp,
  Lurie, and Mavriplis]{slotnick2014cfd}
Jeffrey~P Slotnick, Abdollah Khodadoust, Juan Alonso, David Darmofal, William
  Gropp, Elizabeth Lurie, and Dimitri~J Mavriplis.
\newblock Cfd vision 2030 study: a path to revolutionary computational
  aerosciences.
\newblock Technical report, NASA/CR–2014-218178, 2014.

\bibitem[Kamogawa et~al.(2023)Kamogawa, Tamaki, and
  Kawai]{kamogawa2023ordinary}
Ryo Kamogawa, Yoshiharu Tamaki, and Soshi Kawai.
\newblock Ordinary-differential-equation-based nonequilibrium wall modeling for
  large-eddy simulation.
\newblock \emph{Physical Review Fluids}, 8\penalty0 (6):\penalty0 064605, 2023.

\bibitem[Na and Moin(1998)]{na1998direct}
Y~Na and Parviz Moin.
\newblock Direct numerical simulation of a separated turbulent boundary layer.
\newblock \emph{Journal of Fluid Mechanics}, 374:\penalty0 379--405, 1998.

\bibitem[Goc et~al.(2020{\natexlab{a}})Goc, Bose, and Moin]{gocsubgrid}
Konrad Goc, Sanjeeb.~T. Bose, and Parviz Moin.
\newblock Subgrid-scale modeling sensitivities in wall-modeled large-eddy
  simulations of a high-lift aircraft configuration.
\newblock \emph{Center for Turbulence Research, Annual Research Briefs}, pages
  49--58, 2020{\natexlab{a}}.

\bibitem[Goc et~al.(2020{\natexlab{b}})Goc, Bose, and Moin]{goc2020wall}
Konrad Goc, Sanjeeb Bose, and Parviz Moin.
\newblock Wall-modeled large eddy simulation of an aircraft in landing
  configuration.
\newblock In \emph{AIAA Aviation 2020 Forum}, page 3002, June
  2020{\natexlab{b}}.

\bibitem[Bose and Moin(2014)]{bose2014dynamic}
Sanjeeb~T. Bose and Parviz Moin.
\newblock A dynamic slip boundary condition for wall-modeled large-eddy
  simulation.
\newblock \emph{Physics of Fluids}, 26\penalty0 (1):\penalty0 015104, 2014.

\bibitem[Whitmore et~al.(2021)Whitmore, Bose, and Moin]{whitmorebump}
Michael~P. Whitmore, Sanjeeb~T Bose, and Parviz. Moin.
\newblock Large-eddy simulation of a {G}aussian bump with slip-wall boundary
  conditions.
\newblock \emph{Center for Turbulence Research, Annual Research Briefs}, pages
  45--58, 2021.

\bibitem[Ling et~al.(2022)Ling, Arranz, Williams, Goc, Griffin, and
  Lozano-Dur{\'a}n]{ling2022wall}
Yuenong Ling, Gonzalo Arranz, Emily Williams, Konrad Goc, Kevin Griffin, and
  Adri{\'a}n Lozano-Dur{\'a}n.
\newblock Wall-modeled large-eddy simulation based on building-block flows.
\newblock \emph{arXiv preprint arXiv:2212.05120}, 2022.

\bibitem[Br\`{e}s et~al.(2018)Br\`{e}s, Bose, Emory, Ham, Schmidt, Rigas, and
  Colonius]{bres2018large}
Guillaume~A. Br\`{e}s, Sanjeeb~T. Bose, Michael Emory, Frank~E. Ham, Oliver~T.
  Schmidt, Georgios Rigas, and Tim Colonius.
\newblock Large-eddy simulations of co-annular turbulent jet using a
  {V}oronoi-based mesh generation framework.
\newblock In \emph{2018 AIAA/CEAS Aeroacoustics Conference}, page 3302, 2018.

\bibitem[Agrawal et~al.(2023{\natexlab{b}})Agrawal, Whitmore, Goc, Bose, and
  Moin]{agrawalarb2023_2}
Rahul Agrawal, Michael Whitmore, Konrad Goc, Sanjeeb~T. Bose, and Parviz Moin.
\newblock Reynolds number sensitivities in wall-modeled large-eddy simulation
  of a high-lift aircraft.
\newblock \emph{Center for Turbulence Research, Annual Research Briefs}, pages
  229--244, 2023{\natexlab{b}}.

\bibitem[Tennekes and Lumley(1972)]{tennekes1972first}
Hendrik Tennekes and John~Leask Lumley.
\newblock \emph{A first course in turbulence}.
\newblock MIT press, 1972.

\bibitem[Simpson(1983)]{simpson1983model}
Roger~L Simpson.
\newblock A model for the backflow mean velocity profile.
\newblock \emph{AIAA {J}ournal}, 21\penalty0 (1):\penalty0 142--143, 1983.

\bibitem[Stratford(1959)]{stratford1959prediction}
BS~Stratford.
\newblock The prediction of separation of the turbulent boundary layer.
\newblock \emph{Journal of Fluid Mechanics}, 5\penalty0 (1):\penalty0 1--16,
  1959.

\bibitem[Wei et~al.(2017)Wei, Maciel, and Klewicki]{wei2017integral}
Tie Wei, Yvan Maciel, and Joseph Klewicki.
\newblock Integral analysis of boundary layer flows with pressure gradient.
\newblock \emph{Physical Review Fluids}, 2\penalty0 (9):\penalty0 092601, 2017.

\bibitem[Nickels(2004)]{nickels2004inner}
TB~Nickels.
\newblock Inner scaling for wall-bounded flows subject to large pressure
  gradients.
\newblock \emph{Journal of Fluid Mechanics}, 521:\penalty0 217--239, 2004.

\bibitem[Griffin et~al.(2020)Griffin, Fu, and Moin]{griffinincorporating}
K.~P. Griffin, L.~Fu, and P.~Moin.
\newblock Incorporating non-equilibrium effects in an {ODE}-based wall model.
\newblock \emph{Center for Turbulence Research, Annual Research Briefs}, pages
  73--84, 2020.

\bibitem[Bae and Koumoutsakos(2022)]{bae2022scientific}
H~Jane Bae and Petros Koumoutsakos.
\newblock Scientific multi-agent reinforcement learning for wall-models of
  turbulent flows.
\newblock \emph{Nature Communications}, 13\penalty0 (1):\penalty0 1443, 2022.

\bibitem[Jovic and Driver(1995)]{jovic1995reynolds}
S~Jovic and D~Driver.
\newblock Reynolds number effect on the skin friction in separated flows behind
  a backward-facing step.
\newblock \emph{Experiments in Fluids}, 18:\penalty0 464--467, 1995.

\bibitem[Adams(1984)]{adams1984experiments}
Eric~W. Adams.
\newblock \emph{Experiments on the structure of turbulent reattaching flow}.
\newblock Stanford University, 1984.

\bibitem[Devenport and Sutton(1991)]{devenport1991near}
William~J Devenport and E~Peter Sutton.
\newblock Near-wall behavior of separated and reattaching flows.
\newblock \emph{AIAA journal}, 29\penalty0 (1):\penalty0 25--31, 1991.

\bibitem[Bobke et~al.(2017)Bobke, Vinuesa, {\"O}rl{\"u}, and
  Schlatter]{bobke2017history}
Alexandra Bobke, Ricardo Vinuesa, Ramis {\"O}rl{\"u}, and Philipp Schlatter.
\newblock History effects and near equilibrium in adverse-pressure-gradient
  turbulent boundary layers.
\newblock \emph{Journal of Fluid Mechanics}, 820:\penalty0 667--692, 2017.

\bibitem[Vinuesa et~al.(2018)Vinuesa, Negi, Atzori, Hanifi, Henningson, and
  Schlatter]{vinuesa2018turbulent}
Ricardo Vinuesa, Prabal~Singh Negi, Marco Atzori, Ardeshir Hanifi, Dan~S
  Henningson, and Philipp Schlatter.
\newblock Turbulent boundary layers around wing sections up to ${R}e_c=
  1,000,000$.
\newblock \emph{International Journal of Heat and Fluid Flow}, 72:\penalty0
  86--99, 2018.

\bibitem[Tanarro et~al.(2020)Tanarro, Vinuesa, and
  Schlatter]{tanarro2020effect}
{\'A}lvaro Tanarro, Ricardo Vinuesa, and Philipp Schlatter.
\newblock Effect of adverse pressure gradients on turbulent wing boundary
  layers.
\newblock \emph{Journal of Fluid Mechanics}, 883:\penalty0 A8, 2020.

\bibitem[Lehmkuhl et~al.(2018)Lehmkuhl, Park, Bose, and
  Moin]{lehmkuhl2018large}
O.~Lehmkuhl, G.~I. Park, S.~T. Bose, and P.~Moin.
\newblock Large-eddy simulation of practical aeronautical flows at stall
  conditions.
\newblock \emph{Proceedings of the Summer Program 2018, Center for Turbulence
  Research, Stanford University}, pages 87--96, 2018.

\bibitem[Alber(1971)]{alber1971similar}
Irwin Alber.
\newblock Similar solutions for a family of separated turbulent boundary
  layers.
\newblock In \emph{9th Aerospace Sciences Meeting}, page 203, 1971.

\bibitem[Williams et~al.(2020)Williams, Samuell, Sarwas, Robbins, and
  Ferrante]{williams2020experimental}
Owen Williams, Madeline Samuell, E~Sage Sarwas, Matthew Robbins, and Antonino
  Ferrante.
\newblock Experimental study of a {CFD} validation test case for turbulent
  separated flows.
\newblock In \emph{AIAA Scitech 2020 Forum}, page 0092, 2020.

\bibitem[Gray et~al.(2021)Gray, Gluzman, Thomas, Corke, Lakebrink, and
  Mejia]{gray2021new}
Patrick~D Gray, Igal Gluzman, Flint Thomas, Thomas Corke, Matthew Lakebrink,
  and Kevin Mejia.
\newblock A new validation experiment for smooth-body separation.
\newblock In \emph{AIAA Aviation 2021 Forum}, page 2810, 2021.

\bibitem[Gray et~al.(2022)Gray, Gluzman, Thomas, and
  Corke]{gray2022experimental}
Patrick~D. Gray, Igal Gluzman, Flint~O. Thomas, and Thomas~C. Corke.
\newblock Experimental characterization of smooth body flow separation over
  wall-mounted {G}aussian bump.
\newblock In \emph{AIAA Scitech 2022 Forum}, page 1209, 2022.

\bibitem[Gray et~al.(2022-06)Gray, Gluzman, Thomas, Corke, Lakebrink, and
  Mejia]{gray2022experimentalb}
Patrick~D. Gray, Igal Gluzman, Flint~O. Thomas, Thomas~C. Corke, Matthew~T.
  Lakebrink, and Kevin Mejia.
\newblock Benchmark characterization of separated flow over smooth {Gaussian}
  bump.
\newblock In \emph{AIAA Aviation 2022 Forum}, page 3342. AIAA, 2022-06.
\newblock ISBN 978-1-62410-635-4.
\newblock \doi{10.2514/6.2022-3342}.

\bibitem[Uzun and Malik(2022)]{uzun2021high}
Ali Uzun and Mujeeb~R. Malik.
\newblock High-fidelity simulation of turbulent flow past {G}aussian bump.
\newblock \emph{AIAA Journal}, 60\penalty0 (4):\penalty0 2130--2149, 2022.

\bibitem[Zhou and Bae(2023)]{zhou2023sensitivity}
Di~Zhou and H~Jane Bae.
\newblock Sensitivity analysis of wall-modeled large-eddy simulation for
  separated turbulent flow.
\newblock \emph{preprint, arXiv:2309.13555}, 2023.

\bibitem[Arranz et~al.(2023)Arranz, Ling, and Lozano-Duran]{arranz2023wall}
Gonzalo Arranz, Yuenong Ling, and Adrian Lozano-Duran.
\newblock Wall-modeled les based on building-block flows: Application to the
  gaussian bump.
\newblock In \emph{AIAA Aviation 2023 Forum}, page 3984, 2023.

\bibitem[Agrawal et~al.(2022)Agrawal, Bose, and Moin]{agrawalarb2022}
Rahul Agrawal, Sanjeeb~T. Bose, and Parviz Moin.
\newblock Wall modeled {L}{E}{S} of the {B}oeing speed bump using a
  non-{B}oussinesq modeling framework.
\newblock \emph{Center for Turbulence Research, Annual Research Briefs}, pages
  43--58, 2022.

\bibitem[Agrawal et~al.(2023{\natexlab{c}})Agrawal, Bose, Griffin, and
  Moin]{agrawal2023thwaites}
Rahul Agrawal, Sanjeeb~T. Bose, Kevin~P. Griffin, and Parviz Moin.
\newblock An extension of {T}hwaites method for turbulent boundary layers.
\newblock \emph{preprint, arXiv:2310.16337}, 2023{\natexlab{c}}.

\bibitem[Zhou et~al.(2023)Zhou, Whitmore, Griffin, and Bae]{zhou2023large}
Di~Zhou, Michael~P Whitmore, Kevin~P Griffin, and Hyunji~Jane Bae.
\newblock Large-eddy simulation of flow over boeing gaussian bump using
  multi-agent reinforcement learning wall model.
\newblock In \emph{AIAA Aviation 2023 Forum}, page 3985, 2023.

\bibitem[White(2008)]{white2008fluid}
Frank~M White.
\newblock \emph{Fluid mechanics}.
\newblock The McGraw Hill Companies, 2008.

\bibitem[Bachalo and Johnson(1986)]{bachalo1986transonic}
WD~Bachalo and DA~Johnson.
\newblock Transonic, turbulent boundary-layer separation generated on an
  axisymmetric flow model.
\newblock \emph{AIAA Journal}, 24\penalty0 (3):\penalty0 437--443, 1986.

\bibitem[Lynch et~al.(2023)Lynch, Lance, Miller, Barone, and
  Beresh]{lynch2023experimental}
Kyle~P Lynch, Blake~W Lance, Nathan~E Miller, Matthew~F Barone, and Steven~J
  Beresh.
\newblock Experimental characterization of an axisymmetric transonic separated
  flow for computational fluid dynamics validation.
\newblock \emph{AIAA Journal}, 61\penalty0 (4):\penalty0 1623--1638, 2023.

\bibitem[Jespersen et~al.(2016)Jespersen, Pulliam, and
  Childs]{jespersen2016overflow}
Dennis~C Jespersen, Thomas~H Pulliam, and Marissa~Lynn Childs.
\newblock Overflow turbulence modeling resource validation results.
\newblock Technical report, NASA, NAS-2016-01, 2016.

\bibitem[Spalart et~al.(2017)Spalart, Belyaev, Garbaruk, Shur, Strelets, and
  Travin]{spalart2017large}
Philippe~R Spalart, Kirill~V Belyaev, Andrey~V Garbaruk, Mikhail~L Shur,
  Mikhail~Kh Strelets, and Andrey~K Travin.
\newblock Large-eddy and direct numerical simulations of the bachalo-johnson
  flow with shock-induced separation.
\newblock \emph{Flow, Turbulence and Combustion}, 99\penalty0 (3-4):\penalty0
  865--885, 2017.

\bibitem[Lv et~al.(2021)Lv, Yang, Park, and Ihme]{lv2021discontinuous}
Yu~Lv, Xiang~IA Yang, George~I Park, and Matthias Ihme.
\newblock A discontinuous galerkin method for wall-modeled large-eddy
  simulations.
\newblock \emph{Computers \& Fluids}, 222:\penalty0 104933, 2021.

\bibitem[Horstman and Johnson(1984)]{horstman1984prediction}
CC~Horstman and DA~Johnson.
\newblock Prediction of transonic separated flows.
\newblock \emph{AIAA journal}, 22\penalty0 (7):\penalty0 1001--1003, 1984.

\bibitem[Schatzman and Thomas(2017)]{schatzman2017experimental}
DM~Schatzman and FO~Thomas.
\newblock An experimental investigation of an unsteady adverse pressure
  gradient turbulent boundary layer: embedded shear layer scaling.
\newblock \emph{Journal of Fluid Mechanics}, 815:\penalty0 592--642, 2017.

\bibitem[Knopp et~al.(2021)Knopp, Reuther, Novara, Schanz, Sch{\"u}lein,
  Schr{\"o}der, and K{\"a}hler]{knopp2021experimental}
Tobias Knopp, Nico Reuther, Matteo Novara, Daniel Schanz, Erich Sch{\"u}lein,
  Andreas Schr{\"o}der, and CJ~K{\"a}hler.
\newblock Experimental analysis of the log law at adverse pressure gradient.
\newblock \emph{Journal of Fluid Mechanics}, 918:\penalty0 A17, 2021.

\bibitem[Lozano-Dur\'{a}n and Bae(2019)]{lozano2019characteristic}
Adri\'{a}n Lozano-Dur\'{a}n and Hyunji~Jane Bae.
\newblock Characteristic scales of {T}ownsend's wall-attached eddies.
\newblock \emph{Journal of Fluid Mechanics}, 868:\penalty0 698--725, 2019.

\bibitem[Kim et~al.(1987)Kim, Moin, and Moser]{kim1987turbulence}
John Kim, Parviz Moin, and Robert Moser.
\newblock Turbulence statistics in fully developed channel flow at low
  {R}eynolds number.
\newblock \emph{Journal of Fluid Mechanics}, 177:\penalty0 133--166, 1987.

\end{thebibliography}

\end{document}